\documentclass[11pt,letterpaper]{article}
\usepackage{amsfonts,mathrsfs,natbib,hyperref}

\usepackage{pdflscape}

\usepackage[section]{placeins}
\usepackage{amssymb}
\usepackage{amsmath}
\usepackage{float}
\usepackage[margin=1.0in]{geometry}
\usepackage{color}
\usepackage[dvips]{graphicx}
\usepackage{Here,multirow,lscape}
\usepackage[most]{tcolorbox}
\usetikzlibrary{patterns,shadows}
\usepackage{amsthm}

\newtheorem{asu}{Assumption}
\newcounter{subassumption}[asu]
\renewcommand{\thesubassumption}{(\textbf{\alph{subassumption}})}
\makeatletter
\renewcommand{\p@subassumption}{\theasu}
\makeatother
\newcommand{\subasu}{
	\refstepcounter{subassumption}%
	\thesubassumption~\ignorespaces}
\newtheorem{theorem}{Theorem}[section]

\date{\today}
\newtcolorbox{boxone}{opacityback=0,enhanced jigsaw}

\usepackage{authblk}
\begin{document}

\title{Partially Identified Heterogeneous Treatment Effect with Selection: An Application to Gender Gaps
}
\author[1]{Xiaolin Sun}
\author[1]{Xueyan Zhao}
\author[1]{D.S. Poskitt}
\affil[1]{Monash University \protect}
\maketitle
\begin{abstract}
This paper addresses the sample selection model within the context of the gender gap problem, where even random treatment assignment is affected by selection bias. By offering a robust alternative free from distributional or specification assumptions, we bound the treatment effect under the sample selection model with an exclusion restriction, an assumption whose validity is tested in the literature. This exclusion restriction allows for further segmentation of the population into distinct types based on observed and unobserved characteristics. For each type, we derive the proportions and bound the gender gap accordingly. Notably, trends in type proportions and gender gap bounds reveal an increasing proportion of always-working individuals over time, alongside variations in bounds, including a general decline across time and consistently higher bounds for those in high-potential wage groups. Further analysis, considering additional assumptions, highlights persistent gender gaps for some types, while other types exhibit differing or inconclusive trends. This underscores the necessity of separating individuals by type to understand the heterogeneous nature of the gender gap.

\vspace{.5cm}

\vspace{.5cm}
\end{abstract}

\newpage
\section{Introduction}

In this study, we contribute to the understanding of gender gaps in labor markets through two key avenues. Firstly, we introduce a partial identification method that bounds gender gaps without excessive specification assumptions, maintaining certain literature-based assumptions while relaxing others. Secondly, we move beyond conventional wage and employment status comparisons and delve into subpopulations defined by various traits, including years, covariate groups, and importantly, types of individuals. This approach allows us to uncover heterogeneous treatment effects based on both observable and unobservable characteristics.

A key focus of our method lies in obtaining gender gaps for different types based on these characteristics. For example, we consider types of individuals who may opt to work under certain conditions but not under others, reflecting diverse motivations and social norms. Understanding the gender gap across such types is important as it sheds light on the complexities of individual choices and societal expectations. Even among nonworkers, identifying the gender gap for different types remains important, offering insights into disparities shaped by factors such as family dynamics and personal motivations. By investigating wage gap bounds for different types of individuals under these distinct employment scenarios, influenced by both observable attributes like age and education, and underlying societal norms or motivations, our study provides valuable insights into gender disparities across diverse subpopulations. This comprehensive approach enhances our understanding of treatment effects within varied social contexts and individual circumstances.

Labor force participation and wages serve as vital metrics for assessing the performance of subgroups or types within a population. Labor force participation, representing whether individuals choose to work or not, holds significance as wages are observable only for those who are employed. Consequently, the absence of wage data for non-working individuals is not random but rather stems from their deliberate decisions, highlighting the importance of addressing selection issues in gender gap analyses.

Various approaches exist to tackle the selection problem in gender gap research and related fields. Some methods involve making distributional assumptions and specifying models to estimate parameters in wage equations, followed by a comparison of these parameters between men and women. Others focus on identifying average treatment effects using information from the conditional distribution of selected individuals. Another approach involves considering both working and non-working individuals and employing assumptions to bound the wage distributions for men and women separately. This approach utilizes partial identification techniques, wherein the wage distributions are bounded first, followed by the calculation of differences between these distributions to quantify the gender gap.

We propose another partial identification method within the potential outcome framework to directly assess the effect of gender on wages, bypassing the need to bound wage distributions separately for men and women, whether employed or not. Instead, we bound the Average Treatment Effect (ATE) of gender based on different types, offering a distinct approach from traditional methods reliant on wage equation estimation. Our method does not hinge on specific model specifications or distributional assumptions, unlike methods estimating wage equations, which often require such assumptions. Consequently, rather than estimating the gender gap parameter, our method bounds it, providing a robust check or supplement to existing results. Notably, works by Heckman, such as \cite{Heckman1974} and \cite{Heckman1979}, provide point identification for treatment effects in models with selection. However, their approaches rely on specific models and distribution assumptions. For instance, \cite{Heckman1974} and \cite{Heckman1979} utilize the relationship between the conditional and unconditional distributions of unobserved error terms to point identify parameters, relying on conditions such as the comparison between wages and reservation wages or the inclusion of selection rules in model equations. Despite their contributions, these methods require specific models and distribution assumptions, whereas our approach offers a more flexible and robust alternative. In gender gap studies, \cite{BlauBeller1988} and \cite{Mulligan2008} employ the model from \cite{Heckman1979} to estimate the parameters in the wage equations.

While our method shares similarities with approaches identifying average treatment effects through conditional distribution information from selected samples, we distinguish ourselves by incorporating conditional distributions on observable covariates and the selection, which integrates the underling unobservable covariates. Unlike methods seeking point identification, our estimation method does not involve estimating distributions, conditional expectations, or copula parameters. Utilizing data from the Current Population Survey (CPS) in the United States spanning 1976 to 2013, sourced from \cite{Maasoumi2019}, our findings serve as a complementary analysis to theirs. In contrast, \cite{Fernandez2021} (following \cite{Blundell2007}) analyze data from the Family Expenditure Survey (FES) in the United Kingdom, focusing on the effect of education on log hourly wage with a censored outcome variable and a non-binary selection variable, namely, working hours. Their approach centres on identifying functions such as conditional expectations and distributions of observable variables, employing semiparametric methods like logistic  and quantile regression. They observe a relatively stable effect of education but note an increasing impact of experience on wages in the 1990s.

Expanding upon their methodology, \cite{Fernandez2023} introduce a secondary selection equation into the model to address selection bias based on the distribution of hourly wages among women. Their approach, based on a ``control function" methodology with identification from \cite{Buchinsky1995}, involves semiparametric estimation techniques applied to CPS data spanning 1975 to 2020. They detect positive selection and note evolving selection effects over time for all workers (full-time workers and part-time workers). Throughout the estimation process, they make use of various specification assumptions regarding distributions and apply parametric and distribution regressions. 

Instead of imposing functional form assumptions, \cite{Huber2014ER} identifies the average and quantile treatment effects by adapting inverse probability weighting to the sample selection model. The method relies on the exclusion restriction—using an instrument for selection—to identify the sample selection propensity score. This propensity score is then used as a `control function.' The author applies this method to estimate the returns of high school graduation for females. The instrument, which is the number of young children and its interaction terms, must include a continuous variable for identification purposes, but it is not intended to identify more types in the population.

In contrast, \cite{Maasoumi2019} employ a quantile copula function approach, as introduced by \cite{ArellanoBook2017} and \cite{Arellano2017}, to estimate parameters in quantile wage functions under quantile selection models. This approach utilizes the Frank copula. The copula parameter serves to capture the dependence between error terms within both the wage and selection equations. Implementing this method entails three primary steps under point identification. Notably, their approach necessitates an exclusion restriction for the selection part, and they define the gender gap unconventionally through evaluation functions, focusing solely on full-time workers within their dataset and reporting gender gap measures both in conventional metrics, such as the average, and in an unconventional way.

Our method bears resemblance to partial identification techniques employed in previous studies, such as the approach used by \cite{Blundell2007}, which partially identifies the distributions of log hourly wages for men and women through bounds. Specifically, they focus on bounding the distributions of wages for nonworking individuals based on assumptions about the distribution of workers and nonworkers; the wage distribution of nonworkers is first-order stochastically dominated by the distribution of workers. However, unlike their approach, we treat gender as a treatment and bound the gender gap by types determined by observable and unobservable characteristics. Utilizing data from the Family Expenditure Survey (FES) in the United Kingdom, \cite{Blundell2007} finds that under certain assumptions, including stochastic dominance assumptions, the bounds for changes in the quantile treatment effect of gender at the median suggest a decline for individuals without college experience, while the gender gap decline is less apparent for college-educated individuals. Notably, their approach assumes positive selection for men and women, considering differences in the wage distributions between workers and nonworkers without categorizing individuals into specific types. Thus, while both methods aim to bound treatment effects, our approach offers a nuanced understanding of the gender gap by considering diverse types of individuals.

Our method aligns with the partial identification approach under a sample selection model, building on the foundational work of \cite{Manski1990}, who introduced nonparametric bounds for treatment effects. These bounds initially remain wide without additional assumptions, but \cite{Manski1990} proposed methods to narrow them by incorporating specific assumptions and intersections. Within the framework of sample selection models, economists have developed bounds on ATE for various subpopulations. For instance, \cite{Zhang2003} established upper and lower bounds for the ATE in the ``always observed'' subpopulation, where outcome variables are observed irrespective of treatment status. Similarly, \cite{Lee2009} derived bounds for this subpopulation under monotonicity assumptions. \cite{Zhang2008} extended this work by bounding treatment effects for four subpopulations, including those always observed, only observed when treated, observed when untreated, and never observed individuals. They provided results without relying on monotonicity or stochastic dominance assumptions and contrasted these with results under those assumptions. Additionally, \cite{Blanco2013} began with a sample selection model devoid of distributional and monotonicity assumptions and applied nonparametric bounding methods proposed by \cite{Horowitz2000}. They progressively introduced more assumptions to obtain tighter bounds on the average treatment effects of training programs such as Job Corps on wages, focusing on always-employed individuals and utilizing monotonicity and stochastic dominance assumptions to narrow the bounds further.

Our approach adopts the bounding method introduced in \cite{Lee2009} and \cite{Zhang2008}, employing a nonseparate model framework where gender serves as the ``treatment". Specifically, we measure the gender gap as the disparity between a man's log hourly wage and the potential outcome if he were treated as a woman. Unlike \cite{Lee2009} and \cite{Zhang2008}, we incorporate an exclusion restriction, to distinguish between various population segments beyond the conventional distinctions of ``always observed,'' ``only observed when treated,'' ``observed when untreated,'' and ``never observed'' individuals. This segmentation allows us to delineate types associated with social norms and family roles, defined by both observable and unobservable characteristics. Thus, we are able to bound the gender gap, defined as the difference between the average log hourly wages of women and men, or the ATE for each type, conditional on some observed characteristics. Consequently, the treatment effect becomes heterogeneous across different types, and the introduction of the exclusion restriction enables us to obtain more information from the instrument for selection, bounding the ATE for each value of the instrument. We then intersect these bounds by instrument values, later adjusting them using the CLR method from \cite{Chernozhukov2013}.

Based on the CPS data from \cite{Maasoumi2019}, we draw three key insights. Overall, our findings underscore the significance of accounting for diverse individual types when assessing the gender gap. Firstly, for the dominant type of individuals who always work, we observe rapid declines in the upper bound (UB) of the gender gap until the 1990s, followed by a slower decrease thereafter, with occasional rebounds. Conversely, the decline in the lower bound (LB) is notably slower, and sometimes in later years, the LB is below 0, aligning with \cite{Maasoumi2019}'s results and suggesting that we do not rule out the possibility that the gender gap disappears. Additionally, the UB and LB for high-potential wage workers consistently surpass those for their low-potential counterparts, corroborating the findings in \cite{Maasoumi2019}. Notably, this type always chooses to work under all kinds of conditions with high motivation. The rising proportion of those individuals is mainly due to the fact that more and more women with young children at home choose to work. Also, these results are obtained without assuming any stochastic dominance assumptions, though imposing such assumptions increases the LB, excluding 0 from the bounds while maintaining similar trends.

Secondly, we observe that the bounds for ATE vary significantly across the other types of individuals compared to the relatively stable bounds for those who always work. This variability arises from shifts in the proportions of these types over time. Notably, as the proportion of women opting not to work, particularly those with young children at home, decreases and the proportion of women choosing to work remains stable or increases, the bounds for ATE widen considerably for these types. This increased variability in bounds, reflected in fluctuating LBs and UBs, likely stems from the smaller proportion of individuals in these types. Despite some bounds increasingly excluding 0 in later years, indicating a persistent gender gap, the wide 95\% confidence intervals suggest a potentially larger gender gap or no gender gap for those types.

Thirdly, our analysis sheds light on the issue of selection using our method, yielding results consistent with those of \cite{Maasoumi2019} regarding women. Specifically, our findings indicate a shift from negative to positive selection for women, aligning with previous research. However, our method reveals that this shift occurs earlier in the timeline, with more years exhibiting positive selection. This observation echoes the conclusions drawn by \cite{Fernandez2023}, who also focus on all workers (full-time and part-time workers), despite differences in definitions, methodologies, and datasets. The consistency of these findings underscores the significance of addressing selection issues in understanding gender disparities in the labour market.

The structure of the paper is as follows: Section 2 outlines the basic model, including the assumptions, bounds, estimators, and their asymptotic properties for ATE. Section 3 extends this analysis by introducing additional assumptions and corresponding bounds. In Section 4, we present the findings from our empirical investigation into the gender gap. Finally, Section 5 offers our conclusions. Additional tables and graphs are provided in the Appendix, with further empirical results detailed in the Supplementary Appendix.

\section{the Basic Model}

In this section, we introduce the basic model, focusing on the impact of the binary treatment variable $D$ on the outcome variable $Y$, while considering unobservable types and observable characteristics within a sample selection model. Unlike the endogeneity of treatment in studies like \cite{Imbens1994} and \cite{Angrist1996}, here the endogenous variable is the selection indicator $S$. $X$ represents a vector of observable individual characteristics.\footnote{Initially, we examine the scenario where covariates ($X$) are excluded. We then add $X$ to the analysis by conditioning the assumptions, the conditional probabilities, and the conditional expectations, for instance, on $X$. The analysis then proceeds along the same lines as when $X$ was excluded.} As discussed in \cite{Mellace2011} and \cite{Huber2012}, we first adopt a structural model:
\begin{align*}
  S &= g(D, Z, V) \\
  Y^* &= \psi(D, U) \\
  Y &= SY^*
\end{align*}
Utilizing the potential outcome framework and following \cite{Lee2009} and \cite{Huber2015}, the structural model becomes:
\begin{align*}
 S &= S(1,Z) D + S(0,Z)(1 - D)\\
 Y^* &= Y^*(1) D + Y^*(0)(1 - D)
\end{align*}

In the context of gender gap studies, for instance, in \cite{Maasoumi2019}, $S$ corresponds to employment status, $D$ represents gender, $Z$ denotes the indicator of not having young children (see below), and $Y$ is log hourly wage rate.

The researcher observes ${D, Z, S, Y}$ with $Z$ taken as an instrumental variable, and by inserting $Z$ within the selection function for $S$ we are able to ascertain different behavior patterns and provide insights into different types of individual.

\begin{asu} \label{asu.indep}
The pair $(D, Z)$ is independent of the pair $(U, V)$; denoted as $(D, Z) \perp (U, V)$. If we include covariates $X$ in the discussion, then this assumption is generalised to $(D, Z) \perp (U, V)|X$.
\end{asu}
Assumption \ref{asu.indep} implies that $(S(d,z), Y^*(d))$ and the treatment are independent, or the potential outcomes are independent of the treatment. \cite{Lee2009} and \cite{Huber2015} define the types without the exclusion restriction, using $S(d)$ to define the types and separate individuals into four subpopulations. In our work, we define the types via $S(d,z)$. After that, we bound the average treatment effect for each type, i.e., $E(Y^*(1) - Y^*(0)|\text{Type T})$. 

The exclusion restriction on $Z$ is testable. \cite{Huber2014} propose a testing method under a potential outcome framework, and \cite{Maasoumi2019} utilize this method to test the validity of the exact same instrument as we use in the application section without rejecting the hypothesis about the exclusion restriction of the instrument.

\begin{asu} \label{asu.thres}
$S = g(D, Z, V)$, where $V \in (0,1)$ is continuously distributed and $g$ is weakly nonincreasing in $V$. The function $g$ is  normalized so that V $\sim U(0,1)$.
\end{asu}

Assumption \ref{asu.thres} is a standard assumption found in works such as \cite{Chesher2010} and \cite{Bartalotti2023}. In \cite{Chesher2010}, this assumption applies to discrete outcome variables, implying that the outcome follows a threshold-crossing model. Similarly, in \cite{Bartalotti2023}, where the treatment variable is discrete and there exists an instrument for it, this assumption suggests that the model equation for the treatment involves threshold crossing. In our context, the instrument enters solely via the selection component, indicating that it is only this part of the model that operates as a threshold-crossing process and hence that Assumption \ref{asu.thres} is not restrictive. After imposing Assumption \ref{asu.thres} the selection equation of the structural model becomes
$S = 1[h(D, Z)-V > 0]$,
with threshold-crossing form for $S(d, Z)$, $d\in \{0,1\}$, given by
\begin{equation*}
  S(d, Z) = \begin{cases}
           0, & \mbox{if } V \geq h(d, Z) \\
           1, & \mbox{if } V < h(d, Z)\,.
         \end{cases}
\end{equation*}

In the first step, we aim to determine the proportions of individuals belonging to each type. Subsequently, we use these proportions to generate bounds for the average treatment effect for each type. In gender gap studies, for instance, this process identifies the proportion of individuals who would work if they were women with young children at home. To find the proportions of a certain type, we identify the proportions of individuals who will work when they are subjected to treatment or remain untreated, influenced by a specific value of the instrument. $P_r(S = 1| D = 1, Z = z)$ represents the proportion of individuals observed ($S = 1$) when they receive treatment ($D = 1$) with $Z = z$, while $P_r(S = 1| D = 0, Z = z)$ indicates the proportion of individuals observed when they are untreated ($D = 0$). With $d\in \{0,1\}$,
$$
P_r(S = 1| D = d, Z = z) = P_r(S(d,z) = 1| D = d, Z = z) = P_r(S(d,z) = 1) =  P_r(V < h(d, z))\,.
$$
The first equality derives from the model, the second from Assumption \ref{asu.indep}, and the final one from Assumption \ref{asu.thres}. These equations show that when Assumption \ref{asu.indep} and Assumption \ref{asu.thres} hold the probability $P_r(S = 1| D = d, Z = z)$ is $P_r(V < h(d, z))$, i.e., $h(d, z)$. That is, if we identify the conditional probabilities, we identify the thresholds.

\subsection{Introduction of Different Types}

In the selection framework, we build upon definitions proposed by \cite{Huber2015} based on potential selection outcomes. For instance, individuals with $S(d) = 1$ in \cite{Huber2015} and \cite{Lee2009} are termed ``always observed," while $(S(1) = 1, S(0) = 0)$ are ``compliers," and $(S(1) = 0, S(0) = 1)$ are ``defiers." However, in our framework, with $S(d,z)$ instead of $S(d)$, we need new definitions. Our application in Section \ref{sec:emp} will clarify the importance of these distinctions.

In our gender gap study, $Z$ indicates the presence of young children and is exclusively tied to the selection part of the model. Thus, $Z$ serves as an indicator of an individuals' observability (whether they are employed or not employed). Without additional assumptions, this implies that there are 16 distinct types determined by the values of $S$, $D$, and $Z$, as shown in the first four columns of Table \ref{tbl:PS.DZ.asu}.

The identification challenge with $S(d,z)$ resembles that of $S(d)$, as each individual is observed with only one $S(d,z)$ value, which can be either 0 or 1. Without additional assumptions, multiple types may correspond to one observed case for all possible combinations of $D$ and $Z$ with $S = 1$ (e.g., 8 types in the case of $(S = 1, D = 1, Z = 1)$), and there will not be point identification for the proportions of each type. We will need to bound the proportions of each type, as in \cite{Huber2015}. Here, we introduce the assumptions to define these types and point identify those proportions, as in \cite{Lee2009}.

\begin{table}[!htbp]
\center
\scalebox{0.8}{
\begin{tabular}{c c c c c c}
  \hline
  S(1,0) & S(1,1) & S(0,0) & S(0,1) & Appellation ($S(d,z)$) & Information with assumptions\\ \hline
  1 & 1 & 1 & 1 &  type 1 &  \\
  1 & 1 & 1 & 0 &  type 2 &  \\
  1 & 1 & 0 & 1 &  type 3 &  Violate Assumption \ref{asu.monoZ} (a)\\
  1 & 1 & 0 & 0 &  type 4 &  \\
  1 & 0 & 1 & 1 &  type 5 &  Violate Assumptions \ref{asu.monoD} and \ref{asu.monoZ} (b)\\
  1 & 0 & 1 & 0 &  type 6 &  Violate Assumption \ref{asu.monoZ} (b)\\
  1 & 0 & 0 & 1 &  type 7 &  Violate Assumptions \ref{asu.monoD}, \ref{asu.monoZ} (a) and (b)\\
  1 & 0 & 0 & 0 &  type 8 &  Violate Assumption \ref{asu.monoZ} (b)\\
  0 & 1 & 1 & 1 &  type 9 &  Violate Assumption \ref{asu.monoD}\\
  0 & 1 & 1 & 0 &  type 10 & Violate Assumption \ref{asu.monoD} \\
  0 & 1 & 0 & 1 &  type 11 &  Violate Assumption \ref{asu.monoZ} (a)\\
  0 & 1 & 0 & 0 &  type 12 &  \\
  0 & 0 & 1 & 1 &  type 13 &  Violate Assumption \ref{asu.monoD}\\
  0 & 0 & 1 & 0 &  type 14 &  Violate Assumption \ref{asu.monoD}\\
  0 & 0 & 0 & 1 &  type 15 &  Violate Assumptions
   \ref{asu.monoD} and \ref{asu.monoZ} (a)\\
  0 & 0 & 0 & 0 &  type 16 &  \\
  \hline
\end{tabular}
}
\caption{Principal Strata}
\label{tbl:PS.DZ.asu}
\end{table}

\begin{asu} \label{asu.monoD} (Monotonicity w.r.t D)
$S(1,z) \geq S(0,z)$ with probability 1.
\end{asu}

Assumption \ref{asu.monoD} asserts that the probability of $S(1,z)$ being greater than or equal to $S(0,z)$ holds with certainty for each $Z = z$. This assumption helps rule out situations where potential selection status decreases as a result of the increased treatment. Assumption \ref{asu.monoD} excludes Types 5, 7, 9, 10, 13, 14, and 15, as detailed in Table \ref{tbl:PS.DZ.asu}. For instance, Types 9, 10, 13, and 14, when $Z = 0$, have potential selection values of $S(1,0)$ less than $S(0,0)$, and Types 5, 7, 13, and 15, when $Z = 1$, have $S(1,1)$ less than $S(0,1)$.

\begin{asu} \label{asu.monoZ} (Monotonicity w.r.t Z by treatment status):\\
\subasu \label{asu.monoZ.D0} $S(0,0) \geq S(0,1)$ with probability 1. \\
\subasu \label{asu.monoZ.D1} $S(1,0) \leq S(1,1)$ with probability 1.
\end{asu}

The direction of the inequalities in Assumptions \ref{asu.monoD} and \ref{asu.monoZ} will vary depending on the specific empirical context. For our example, where $D = 0$ denotes females, Assumption \ref{asu.monoZ.D0} implies that women will not work when they have young children if they are not working when they do not have young children. Similarly, Assumption \ref{asu.monoZ.D1} suggests that for males, having young children at home will work if they work when they do not have young children. Under Assumptions \ref{asu.monoZ.D0} and \ref{asu.monoZ.D1}, certain types (3, 5, 6, 7, 8, 11, and 15) are excluded, as shown in Table \ref{tbl:PS.DZ.asu}.

Assumptions \ref{asu.monoD} and \ref{asu.monoZ} are not testable, but there are corresponding necessary conditions for them. In our empirical application, Assumptions \ref{asu.monoD} aligns with situations where the probability of $S = 1$, given $D = 1$ and $Z = z$, is greater than or equal to the probability when $D = 0$ and $Z = z$. That is, $P_r(S = 1|D = 1, Z = z) \geq P_r(S = 1|D = 0, Z = z)$ for $z \in \{ 0, 1 \}$. Assumption \ref{asu.monoZ.D0} aligns with the data when $P_r(S = 1|D = 0, Z = 0) \geq P_r(S = 1|D = 0, Z = 1)$. Assumption \ref{asu.monoZ.D1} is consistent with the data when $P_r(S = 1|D = 1, Z = 1) \geq P_r(S = 1|D = 1, Z = 0)$. However, it is important to note that these inequalities are necessary but not sufficient conditions for Assumptions \ref{asu.monoD} and \ref{asu.monoZ} to hold; that is, if the inequalities are satisfied in the empirical application it does not mean that the monotonicity assumptions are valid.

\begin{figure}[!htbp]
\caption{Illustration for the types under Assumptions \ref{asu.indep}-\ref{asu.monoZ}}
\label{fig.h.DZ}
  \centering
  \includegraphics[width=15cm]{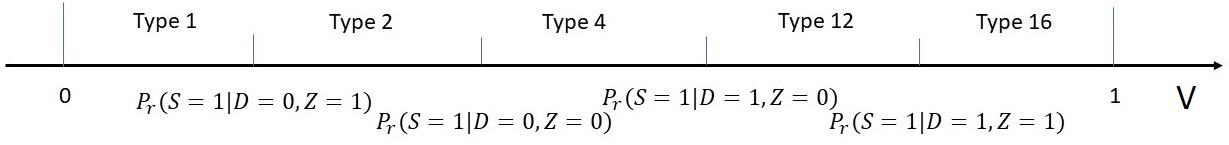}
\end{figure}

\begin{table}[!htbp]
\footnotesize
\begin{center}
\begin{tabular}{c c c c c l}
  \hline
  Strata ($S(d,z)$) & S(1,1) & S(1,0) & S(0,1) & S(0,0) &   \\ \hline
  Type 1 & 1 & 1 & 1 & 1 &   AO w.r.t Z when D is 0 or 1 \\ 
   EEEE &  &    &    &     &   AO w.r.t D when Z = 0 or 1\\ 
  Type 2& 1 & 1 & 0 & 1 &  AO w.r.t Z when D = 1 and D for Z when D = 0\\ 
   EENE &   &    &    &     & C w.r.t D when Z = 1 and AO when Z = 0\\ 
  Type 4& 1 & 1 & 0 & 0 &  AO w.r.t Z when D = 1 and NO when D = 0\\ 
  EENN &  &   &   &        &   C w.r.t D when Z = 1 and C when Z = 0\\ 
  Type 12& 1 & 0 & 0 & 0 &   C w.r.t Z when D = 1 and NO when D = 0\\ 
  ENNN &   &    &    &     &   C w.r.t D when Z = 1 and NO when Z = 0\\ 
 Type 16 &  0 & 0 & 0 & 0 &   NO w.r.t Z when D = 1 and NO when D = 0\\
  NNNN &   &    &    &     &   NO w.r.t D when Z = 1 and NO when Z = 0\\
  \hline
\end{tabular}
\caption{Principal Strata under Assumptions \ref{asu.indep}-\ref{asu.monoZ} with Explanation Using Always Observed (AO), Compliers (C), Defiers (D), and Never Observed (NO) Notations} \label{table:types}
\end{center}
\end{table}

\begin{table}[!htbp]
\center
\begin{tabular}{c c| c c | c c| c c}
  \hline
   \multicolumn{4}{c}{$Z = 1$}  &\multicolumn{4}{c}{$Z = 0$} \\ \hline
   \multicolumn{4}{c}{$D$}  &\multicolumn{4}{c}{$D$} \\
   &  & 0 & 1 &   &   & 0 & 1 \\ \hline
 S & 0 &  Types:  2, 4, 12, 16 & Type: 16  & S & 0 & Types: 4, 12, 16 & Types: 12, 16 \\
  & 1 & Type: 1 & Types: 1, 2, 4, 12 &  & 1 & Types:  1, 2 & Types:  1, 2, 4 \\
  \hline
\end{tabular}
\caption{Types and Observed Sets under Assumptions \ref{asu.indep}-\ref{asu.monoZ}}
\label{TBL:PS.OBSE}
\end{table}

There are five distinct types defined under Assumptions \ref{asu.indep}-\ref{asu.monoZ}, as shown in the final column of Table \ref{tbl:PS.DZ.asu} and visually represented in Figure \ref{fig.h.DZ}. In Table \ref{table:types}, we list the five types or strata according to the value of $S(d,z)$. We introduce another kind of notation for each type. Specifically, we use `E' to denote the condition $S(d, z) = 1$ and `N' to denote $S(d, z) = 0$. For example, the notation `EEEE' represents the stratum consisting of individuals always observed under all combinations of $d \in \{0,1\}$ and $z \in \{0,1\}$, characterized by $\{S(1,1) = S(1,0) = S(0,1) = S(0,0) = 1\}$. Based on this table, we construct Table \ref{TBL:PS.OBSE} to illustrate the selection outcomes corresponding to each type under the case $(D = d, Z = z)$ with $d \in {0,1}$ and $z \in {0,1}$. Table \ref{TBL:PS.OBSE} provides insight into the relationship between those observed cases and the five types.

Let us focus on what each type signifies. Women belonging to Type 2 (EENE stratum) will engage in work if they do not have young children (those aged less than 5 years). On the other hand, women of Type 4 (EENN stratum) will refrain from work regardless of whether they have young children or not. Interestingly, if we were to consider males in Type 2 and Type 4, they would be associated with employment. Conversely, women categorized as Type 12 (ENNN stratum) will abstain from work, regardless of whether they have young children in their care. However, males of the same type would only choose to work when they have young children to look after. From this discussion and Table \ref{TBL:PS.OBSE}, we see that among those who do not work, there are different types. This classification system provides insight into the behaviour of individuals of different types in various scenarios, shedding light on how various factors influence their decisions about work.


\subsection{Identification of the Population Proportions}

Let us now shift our focus to the proportions of each type in the population conditional on $D$ and $Z$. It is important to remember that $V$ represents the error term within the selection equation, effectively influencing the assignment of types. This is shown in Figure \ref{fig.h.DZ}. The unobservable $V$ and the observable characteristics define the types. The relationship between $U$ and $V$ could be quite complex. Under the basic model, we do not assume any assumptions between $U$ and $V$.

Assumption \ref{asu.indep} ensures that $D, Z$ and $V, U$ are independent. Consequently, the proportion of each type conditional on the treatment ($D$) and the instrument ($Z$) is the same as the unconditional proportion of each type in the whole population. In other words, in Table \ref{TBL:PS.OBSE}, the proportion of each type within a specific case $(D = d, Z = z)$ when conditioned on treatment and the instrument is the same across all combinations of $D, Z$. The intuition is the same as in \cite{Lee2009}. For instance, within the cell where $(S = 1, D = 0, Z = 1)$ in Table \ref{TBL:PS.OBSE}, Type 1 prevails and all individuals within the cell are of Type 1. Thus, in cases where $(S = 1, D = 0, Z = 1)$ an individuals type is known and the unconditional proportion of Type 1 ($\pi_{T1}$) is defined as $\pi_{T1} = P_r(S =1| D = 0, Z = 1)$.\footnote{$
  P_r(S =1| D = 0, Z = 1) = P_r(S(0,1) =1| D = 0, Z = 1) = P_r(S(0,1) =1) = P_r(S(0,1) =1, S(0,0) =1,S(1,1) =1,S(1,0) =1) =  P_r(\text{Type 1}) = \pi_{T1}$. The first equality is from the model setting, the second one is from Assumption \ref{asu.indep}, the third is from all of the four Assumptions so far, and the fourth one is from the definition of Type 1.} 

Moving to the cell characterized by $(S = 1, D = 0, Z = 0)$  in Table \ref{TBL:PS.OBSE}, we find individuals belonging to Type 1 or Type 2. Given our knowledge of the unconditional proportion of Type 1, we can determine the proportion of Type 1 within this group as follows: $\frac{\pi_{T1}}{\pi_{T1}+ \pi_{T2}}= \frac{P_r(S =1| D = 0, Z = 1)}{P_r(S =1| D = 0, Z = 0)}$. \footnote{$P_r(S =1| D = 0, Z = 0) = P_r(S(0,0) =1| D = 0, Z = 0) = P_r(S(0,0) =1) = P_r(S(0,0) =1, S(0,1) =1) +  P_r(S(0,0) =1, S(0,1) =0) = P_r(\text{Type 1}) + P_r(\text{Type 2}) = \pi_{T1}+ \pi_{T2}$. Hence, $\frac{P_r(S =1| D = 0, Z = 1)}{P_r(S =1| D = 0, Z = 0)} = \frac{\pi_{T1}}{\pi_{T1}+ \pi_{T2}}$ under the four assumptions.} Simultaneously, we can calculate the proportion of Type 2 within the same group using the following expression: $\frac{\pi_{T2}}{\pi_{T1}+ \pi_{T2}} = \frac{P_r(S =1| D = 0, Z = 0) - P_r(S =1| D = 0, Z = 1)}{P_r(S =1| D = 0, Z = 0)}$ \footnote{$P_r(S =1| D = 0, Z = 0) - P_r(S =1| D = 0, Z = 1) = \pi_{T1}+ \pi_{T2} - \pi_{T1} =  \pi_{T2}$}. These calculations utilize the unconditional proportion of Type 2 ($\pi_{T2}$), which is determined as the difference between the probability of $S = 1$ when $D = 0$ and $Z = 0$ and the probability when $D = 0$ and $Z = 1$: $\pi_{T2} = P_r(S =1| D = 0, Z = 0) - P_r(S =1| D = 0, Z = 1)$.

Next, in the cell $(S = 1, D = 1, Z = 0)$ we observe the presence of three different types: Types 1, 2, and 4. All three types share the characteristic of having $S(1,0) = 1$. We can ascertain the unconditional proportion of Type 4 ($\pi_{T4}$) as follows: $\pi_{T4} = P_r(S =1| D = 1, Z = 0) - P_r(S =1| D = 0, Z = 0)$ \footnote{
$P_r(S =1| D = 1, Z = 0) - P_r(S =1| D = 0, Z = 0) = P_r(S(1,0) =1) - P_r(S(0,0) =1) = P_r(S(1,0) =1, S(0,0) =1)+ P_r(S(1,0) =1, S(0,0) =0) - P_r(S(1,0) =1, S(0,0) =1) = P_r(S(1,0) =1, S(0,0) =0) =  P_r(\text{Type 4}) = \pi_{T4}$}. Consequently, the proportion of Type 4 within this cell can be calculated as: $\frac{\pi_{T4}}{\pi_{T1}+ \pi_{T2}+ \pi_{T4}}$. Meanwhile, the proportion of Type 1 in the same cell is $\frac{\pi_{T1}}{\pi_{T1}+ \pi_{T2}+ \pi_{T4}}$, and that of Type 2 is $\frac{\pi_{T2}}{\pi_{T1}+ \pi_{T2}+ \pi_{T4}}$.

Finally, the probability of Type 12 in the cell characterized by $D = 1, Z = 1, S = 1$ can be determined using the following expression: $\frac{\pi_{T12}}{\pi_{T1}+ \pi_{T2}+ \pi_{T4}+\pi_{T12}}$ with $\pi_{T12} = P_r(S =1| D = 1, Z = 1) - P_r(S =1| D = 1, Z = 0)$
\footnote{ $P_r(S =1| D = 1, Z = 1) - P_r(S =1| D = 1, Z = 0) = P_r(S(1,1) =1| D = 1, Z = 1) - P_r(S(1,0) =1| D = 1, Z = 0) = P_r(S(1,1) =1) - P_r(S(1,0) =1) = P_r(S(1,1) =1, S(1,0) =1) + P_r(S(1,1) =1, S(1,0) =0)- P_r(S(1,0) =1,S(1,1) =1) = P_r(S(1,1) =1, S(1,0) =0) =  P_r(\text{Type 12}) = \pi_{T12}$}.
The proportion of Type 1 within the same cell is determined as: $\frac{\pi_{T1}}{\pi_{T1}+ \pi_{T2}+ \pi_{T4}+\pi_{T12}}$. The proportion of Type 2 can be calculated as: $\frac{\pi_{T2}}{\pi_{T1}+ \pi_{T2}+ \pi_{T4}+\pi_{T12}}$, and the proportion of Type 4 within this group can be determined as: $\frac{\pi_{T6}}{\pi_{T1}+ \pi_{T2}+ \pi_{T4}+\pi_{T12}}$.

This analysis enables us to discern the proportions of different types within various observed cases. In the context of the gender gap study, it sheds light on the proportions of each type. With data spanning multiple years, it reveals how the types evolve over time within the gender gap study. Moreover, by segmenting the data set into different covariate groups for each year, it demonstrates how the covariate groups and types interact. Thus, it contributes to a deeper understanding of the distribution of these types across different scenarios.

\subsection{Derivation of Bounds}

In this subsection, we discuss the bounds for the ATE for all types within the population: Types 1, 2, 4, 12, and 16. These bounds are established under Assumptions \ref{asu.indep}-\ref{asu.monoZ}, collectively referred to as the basic assumptions. Under these assumptions, we establish the widest possible bounds, representing the most conservative scenario. The ATE for Type T is expressed as $E(Y^*(1) - Y^*(0)|\text{Type T})$. For Type 16, without additional assumptions, the ATE bounds extend across the entire real number line $(-\infty, \infty)$. However, by assuming that potential outcomes for all types are bounded within the interval $[Y^{LB}, Y^{UB}]$, denoted as $Y^*(1)$ and $Y^*(0)$, the ATE bounds narrow down to $[Y^{LB} - Y^{UB}, Y^{UB} - Y^{LB}]$.

\subsubsection{Bounds for \texorpdfstring{$E(Y^*(1)|\text{Type T})$}{E(Y(1)|Type T)}}

As discussed earlier, we begin by deriving the sharp bounds for potential outcomes when treated by the 2 values of the instrument. The proportions of each type within cells $S = 1, D = 1, Z = 1$ and $S = 1, D = 1, Z = 0$ are calculated as $\frac{\pi_{T.}}{\pi_{T1}+ \pi_{T2}+ \pi_{T4}+\pi_{T12}}$ and $\frac{\pi_{T.}}{\pi_{T1}+ \pi_{T2}+ \pi_{T4}}$, respectively, under Assumptions \ref{asu.indep} to \ref{asu.monoZ}. Next, we follow the same logic as in Proposition 1a in \cite{Lee2009} and Section 3 of \cite{Huber2015} to find the sharp upper and lower bounds for each value of $Z$. This is standard procedure in sample selection models. Hence, we have two sets of upper and lower bounds for the same parameter.

When $Z = 1$, we have the upper and lower bounds for $E(Y^*(1)|\text{Type T})$. The upper and lower bounds are in the following: $$UB = E[Y|D = 1, S = 1, Y \geq Y_{1 - q_T}, Z = 1],$$ $$LB= E[Y|D = 1, S = 1, Y \leq Y_{q_T}, Z = 1]$$ Here, $q_T$ represents the proportion of Type T in that cell, such as $\frac{\pi_{T1}}{\pi_{T1}+ \pi_{T2}+ \pi_{T4} + \pi_{T12}}$ for Type 1. The bounds for Types 2, 4, and 12 in the same cell are calculated similarly, with the only difference being in the proportions used for each type.\footnote{We show the relations between observable distributions and unobservable distributions for types and proportions in Appendix}

When $Z = 0$, under Assumption \ref{asu.indep}, \ref{asu.thres}, \ref{asu.monoD}, and \ref{asu.monoZ}, the upper and lower bounds for $E(Y^*(1)|\text{Type T})$ are computed as: $$UB = E[Y|D = 1, S = 1, Y \geq Y_{1 - q_T}, Z = 0]$$
$$LB=E[Y|D = 1, S = 1, Y \leq Y_{q_T}, Z = 0]$$
These bounds for Types 1, 2, and 4 are based on the proportions of each type in the respective case.

Since $Y^*(1)$ does not depend on Z, the bounds when $Z = 1$ and $Z = 0$ provide more information to identify the same parameter. As a result, for Types 1, 2, and 4, we take the intersection of these two sets of bounds in the second step, further refining the ATE bounds. This is the same idea of intersection as in \cite{Chesher2010}, \cite{Huber2015}, and \cite{Huber2017}.

\subsubsection{Bounds for \texorpdfstring{$E(Y^*(0)|\text{Type T})$}{E(Y(0)|Type T)} and ATE}

In this subsection, we calculate the average potential outcomes when untreated for Type 1 and Type 2. When $D = 0$ and $Z = 1$, the observed ($S = 1$) type is Type 1. In our empirical application, this means that females with young children (less than 5 years old) who are employed are of Type 1. Recall that in the cell $(S = 1, D = 0, Z = 1)$ of Table \ref{TBL:PS.OBSE}, there is only Type 1. Thus, $E(Y|S = 1, D = 0, Z = 1)$ is the average potential outcome when untreated for Type 1, i.e., $E(Y^*(0)|\text{Type 1})$. \footnote{$ E(Y|D = 0, S = 1, Z = 1) = E(Y^*(0)|D = 0, S(0,1) = 1, Z = 1) = E(Y^*(0)|S(0,1) = 1) = E(Y^*(0)|\text{Type 1})$}.

Next, we identify $E(Y^*(0)|\text{Type 2})$. Notice that Type 1 and Type 2 are inside the cell $(S = 1, D = 0, Z = 0)$ of Table \ref{TBL:PS.OBSE}. With the information we have, we point identify the average potential outcome when untreated for Type 2. Indeed, we point identify the $E(Y^*(0)|\text{Type 1})$, the proportions of Type 1 and Type 2 in the cell ($\frac{\pi_{T1}}{\pi_{T1}+ \pi_{T2}}$ and $\frac{\pi_{T2}}{\pi_{T1}+ \pi_{T2}}$), and the overall average potential outcome when untreated for Type 1 and Type 2, $E(Y|S = 1, D = 0, Z = 0)$. The point identification for the untreated potential outcome of Type 2 is achieved. This is shown in the following. The proof is in the Appendix.
$$E(Y|S = 1, D = 0, Z = 0) = E(Y^*(0)|\text{Type 1}) * \frac{\pi_{T1}}{\pi_{T1}+ \pi_{T2}} + E(Y^*(0)| \text{Type 2})*\frac{\pi_{T2}}{\pi_{T1}+ \pi_{T2}}$$

Notice that individuals of Types 4 and 12, when untreated, are not observed for any value of $Z$. That is, in Table \ref{TBL:PS.OBSE} or Table \ref{table:types}, $S (0, z)$ is 0 for these two types. Hence, we do not obtain any information from the observed data for $E(Y^*(0)|\text{Type T})$ with $T \in \{4, 12\}$. Without any further assumption, we use theoretical bounds for those values: $E(Y^*(0)|\text{Type T}) \in [Y^{LB}, Y^{UB}]$.

To summarize, the bounds for the ATE for Types 1 and 2 are calculated as follows: $$UB_{T} = \min_{z} E[Y|D = 1, S = 1, Y \geq Y_{1 - q_{T_z}}, Z = z]-E(Y^*(0)|\text{Type T})$$
$$LB_{T}=  \max_{z} E[Y|D = 1, S = 1, Y \leq Y_{q_{T_z}}, Z = z]-E(Y^*(0)|\text{Type T})$$

For Type 4, the ATE bounds are: $$UB_{T4} = \min_{z} E[Y|D = 1, S = 1, Y \geq Y_{1 - q_{T4_0}}, Z = z]-Y^{LB}$$
$$LB_{T4}=  \max_{z} E[Y|D = 1, S = 1, Y \leq Y_{q_{T4_0}}, Z = z]-Y^{UB}$$

And for Type 12: $$UB_{T12}=  E[Y|D = 1, S = 1, Y \geq Y_{1 - q_{T12_0}}, Z = 1]-Y^{LB}$$
$$LB_{T12}=  E[Y|D = 1, S = 1, Y \leq Y_{q_{T12_0}}, Z = 1]-Y^{UB}$$

The bounds are sharp under Assumptions \ref{asu.indep} to \ref{asu.monoZ}. That is, they are the tightest bounds that we get under the assumptions. The detailed discussion on the sharpness of the bounds is in Proposition 1a. from \cite{Lee2009}.

\subsection{Bounds with Covariates}

In the previous discussion, we temporarily excluded the consideration of covariates in our model. Here, we delve into the explicit inclusion of covariates $X$ or individual characteristics. It is straightforward. That is, we apply the discussion, proofs, and results with respect to each value of the covariate in $X$, as in \cite{Lee2009}. This is a similar method to working with covariates, as in \cite{Angrist1996}. That is, if education level is one of the variables inside covariates $X$, the discussion is applied with respect to each education level. In the gender gap study, the covariates include many individual characteristics. We will discuss this in Section \ref{sec:emp}.

To integrate $X$ into our analysis, we introduce an amended set of assumptions. For example, Assumption \ref{asu.indep} now takes the form: $D$ and $Z$ are independent of $U$ and $V$ given $X$. Similarly, monotonicity assumptions are based on each value of each covariate in $X$. We directly update the assumption versions. The revised version of Assumption \ref{asu.indep} is a conditional independence assumption discussed in \cite{Holland1986} and is a direct extension of the confoundedness assumption from \cite{Imbens2004Review}.

In the gender gap study, we employ this revised set of assumptions, recognizing the significance of covariates in influencing both outcomes and selection. These covariates, along with their polynomials, are also accounted for in \cite{Maasoumi2019}, where we adopt the same covariates and polynomial specifications.

The bounds for the ATE for Types 1 and 2 are calculated as: $$UB_{T} = \min_{z} E[Y|D = 1, S = 1, Y \geq Y_{1 - q_{T_z}}, Z = z, X = x]-E(Y^*(0)|\text{Type T}, X = x)$$
$$LB_{T}=  \max_{z} E[Y|D = 1, S = 1, Y \leq Y_{q_{T_z}}, Z = z, X = x]-E(Y^*(0)|\text{Type T}, X = x)$$

For Type 4, the ATE bounds are: $$UB_{T4} = \min_{z} E[Y|D = 1, S = 1, Y \geq Y_{1 - q_{T4_z}}, Z = z, X = x]-Y^{LB}$$
$$LB_{T4}=  \max_{z} E[Y|D = 1, S = 1, Y \leq Y_{q_{T4_z}}, Z = z, X = x]-Y^{UB}$$

And for Type 12: $$UB_{T12} =  E[Y|D = 1, S = 1, Y \geq Y_{1 - q_{T12_1}}, Z = 1, X = x]-Y^{LB}$$
$$LB_{T12}=  E[Y|D = 1, S = 1, Y \leq Y_{q_{T12_1}}, Z = 1, X = x]-Y^{UB}$$

The bounds for the various types are sharp, following the approach outlined in \cite{Lee2009} and \cite{Blanco2013}. We employ the covariates to tighten these bounds, as they possess predictive power for $Y$. By conditioning on the covariates, we group individuals with similar predicted values of $Y$, thereby potentially reducing the variability in $Y$. Given that our bounding method relies on the values of $Y$, the intervals for treatment effects based on covariate groups are consequently narrower. When examining the overall effect of the ATE across all $X$ groups, we compute the average LB and UB across these groups based on the method proposed by \cite{Lee2009} for calculating `total' in Table 5.

\section{Estimation and Inference} \label{sec:estimator}

In this subsection, we compute the sample versions of the population parameters outlined in the previous subsection, following the methodology presented in \cite{Lee2009}. For illustrative purposes, we simplify by disregarding the covariates in the bounds. Additionally, we streamline the illustration by only furnishing estimates for Type 1 and Type 2 when $Z = 1$. The estimators for Type 4 and Type 12 follow a similar approach. The estimators for Type 1 when $Z = 1$ are:
\begin{align}\label{equ:estT1}
  \widehat{UB_{T1}} &= \frac{\sum Y \cdot S \cdot D \cdot Z \cdot 1[Y \geq \widehat{y_{1-\widehat{q}}}]}{\sum S \cdot D \cdot Z \cdot 1[Y \geq \widehat{y_{1-\widehat{q}}}]} -  \widehat{Y_{0,T1}}\\
  \widehat{LB_{T1}}  &= \frac{\sum Y \cdot S \cdot D \cdot Z \cdot 1[Y \leq \widehat{y_{\widehat{q}}}]}{\sum S \cdot D \cdot Z \cdot 1[Y \leq \widehat{y_{\widehat{q}}}]} -  \widehat{Y_{0,T1}} \nonumber \\
  \widehat{y_p} &= min \left\{y: \frac{S \cdot D \cdot Z \cdot 1[Y \leq y]}{S \cdot D \cdot Z} \geq p \right\} \nonumber \\
  \widehat{q} &= \frac{(\sum S \cdot (1-D) \cdot Z)/(\sum (1-D) \cdot Z)}{(\sum S \cdot D \cdot Z)/(\sum D \cdot Z)} \nonumber\\
  \widehat{Y_{0,T1}} &= \frac{\sum Y \cdot S \cdot (1-D)  \cdot Z }{\sum S \cdot (1-D) \cdot Z} \nonumber
\end{align}
The initial component of $\widehat{UB_{T1}}$ contains the upper bound for $E(Y^*(1)|\text{Type 1})$, while the corresponding segment of $\widehat{LB_{T1}}$ is the lower bound for $E(Y^*(1)|\text{Type 1})$. $\widehat{y_p} $ is the quantile, and $ \widehat{q}$ is the proportion of Type 1 in the case of $(S = 1, D = 1, Z = 1)$ in Table \ref{TBL:PS.OBSE}. The numerator of $\widehat{q}$ serves as an estimator for $\pi_{T1}$ and maintains consistency across different values of $Z$. The denominator of $\widehat{q}$, however, changes with the values of $Z$. The ultimate equation furnishes the estimate for $E(Y^*(0)|\text{Type 1})$. When computing the bounds for $Z = 0$, we substitute $Z$ with $(1-Z)$ in the equations above, with the exception of the numerator of $\widehat{q}$ and the final equation.

For Type 2, the estimators are similar except for the last two equations, that is, the estimator for the proportion of Type 2 for each value of $Z$ and the one for $E(Y^*(0)|\text{Type 2})$. For this type,
\begin{align}\label{equ:estT2}
 \widehat{q} &= \frac{\widehat{\pi_{T2}}}{(\sum S \cdot D \cdot Z)/(\sum D \cdot Z)} \nonumber\\
 \widehat{Y_{0,T2}}&= \left[\frac{\sum Y \cdot S \cdot (1-D)  \cdot (1-Z)}{\sum S \cdot (1-D) \cdot (1-Z)} * (\widehat{\pi_{T1}}  + \widehat{\pi_{T2}})-  \widehat{Y_{0,T1}} * \widehat{\pi_{T1}} \right]/\widehat{\pi_{T2}}  \\
  \widehat{\pi_{T1}}+\widehat{\pi_{T2}} &= \frac{\sum S \cdot (1-D) \cdot (1-Z)}{ \sum (1-D) \cdot (1-Z) } \nonumber \\
  \widehat{\pi_{T1}} &= \frac{\sum S \cdot (1-D) \cdot Z}{\sum (1-D) \cdot Z} \nonumber
\end{align}
Note that the denominator of $\widehat{Y_{0,T2}}$ is $\widehat{\pi_{T2}}$. In our gender gap study, this proportion is a positive number, which is a necessary condition for Assumptions \ref{asu.monoD} and \ref{asu.monoZ} to be valid.

For Types 4 and 12, the estimates $E(Y^*(0)|\text{Type T})$ under the basic assumptions are in the theoretical bounds $[Y^{LB}, Y^{UB}]$. We follow the steps in \cite{Huber2015} using the minimum and maximum values from the dataset for each year.

\subsection{Asymptotic Properties}

We derive the asymptotic properties, including consistency and asymptotic normality, for the estimators in Section \ref{sec:estimator} by employing the GMM method proposed in Propositions 2 and 3 of \cite{Lee2009}. This approach is suitable as the estimators represent solutions to just-identified GMM problems, as outlined in \cite{Newey1994}. Specifically, the parameters of interest are included within moments, resulting in a just-identified case where the number of parameters to estimate equals the number of moments. Using these moments, we construct sample analogs and subsequently derive GMM-type estimators, as detailed in Section \ref{sec:estimator}. Drawing from the methodologies outlined in \cite{Newey1994} and \cite{Lee2009}, we establish the asymptotic properties for these estimators. Our approach aligns with the assumptions proposed by \cite{Lee2009}, with additional considerations incorporating the instrument variable as an extra condition for each type.

\begin{theorem} (Consistency and Asymptotic Normality) \label{theoremCAN}
$Y(1)$, $Y(0)$ are bounded. With Assumptions \ref{asu.indep} - \ref{asu.monoZ}, and extra assumptions $0 < E[S|D = d, Z = z] < 1$ for $d, z \in \{0, 1\}$, $ \widehat{UB_T} \rightarrow^{p} UB_T$ and $\widehat{LB_T}\rightarrow^{p} LB_T$, as well as\\
 $\sqrt{n}(\widehat{UB_T} - UB_T) \rightarrow^d N(0, V^{UB}_T + V^C_T)$ and $\sqrt{n}(\widehat{LB_T} - LB_T) \rightarrow^d N(0, V^{LB}_T + V^C_T)$ with $V^{LB}_T$ and $V^C_T$ defined for each type and $Z$ in the Appendix.
\end{theorem}

Given that $0 < E[S|D = d, Z = z] < 1$ for $d, z \in \{0, 1\}$ this indicates that the conditional probabilities are constrained to be between 0 and 1. With Assumptions \ref{asu.indep} through \ref{asu.monoZ}, it follows that the proportions of types also fall within the range of 0 to 1. The proof for Theorem \ref{theoremCAN} is provided in the appendix. In the appendix, we also provided the covariance of $\widehat{UB}_T - UB_T$ and $\widehat{LB}_T - LB_T$ when $Z = 1$ and $Z = 0$ and the covariance between $V^C_T$ of Type 1 and $V^C_T$ of Type 2. Those covariances are needed to calculate the CLR bounds from \cite{Chernozhukov2013}. CLR bounds are necessary, because the intersection bounds contain the min and max operators, creating biased UBs and LBs. In Section \ref{sec:emp}, we obtain half-median unbiased estimates of UBs and LBs, as well as confidence intervals using the method from \cite{Chernozhukov2013}.

\section{Further Assumptions and Tighter Bounds}
\subsection{Mean Dominance Assumption}

In this section, we introduce further assumptions to tighten the bounds. The utilization of stochastic dominance assumptions in sample selection frameworks has been demonstrated by previous researchers. See, inter alia, \cite{Zhang2003}, \cite{Zhang2008}, \cite{Blanco2013}, \cite{Huber2015}, and \cite{Bartalotti2023}. For instance, \cite{Bartalotti2023} have employed stochastic assumptions to refine the bounds for individuals always observed. They assume that the distribution of potential outcomes when treated for the always observed subpopulation exhibits first-order stochastic dominance over the distributions of these potential outcomes for compliers, defiers, or those who are never observed. Notably, \cite{Bartalotti2023} introduce this assumption when an instrument is used for the treatment, so there will be extra conditions under their setting. \cite{Blanco2013} and \cite{Huber2017}, on the other hand, rely on a mean dominance assumption. We adopt similar assumptions to \cite{Blanco2013} and \cite{Huber2017} because we are focused on the mean effects for each type.

We apply Assumptions \ref{asu.FOSD.Y0} and \ref{asu.FOSD.Y1} in our gender gap application where the treatment is exogenous and the instrument pertains to the selection process. \footnote{It is important to note that the direction of the inequality is based on the context or the specific application. In a different application, the direction of the inequality may change. This is similar to Assumptions \ref{asu.monoD} and \ref{asu.monoZ}.}

\begin{asu} (Mean Dominance Assumption) \\
\subasu \label{asu.FOSD.Y0}   $E(Y^*(0)| \text{Type j}) \geq E(Y^*(0)| \text{Type k})$ with $j \in \{1, 2\}$ and $k \in \{4, 12\}$\\
\subasu  \label{asu.FOSD.Y1}  $E(Y^*(1)| \text{Type 1}) \geq E(Y^*(1)| \text{Type 2}) \geq E(Y^*(1)| \text{Type 4}) \geq E(Y^*(1)| \text{Type 12})$
\end{asu}

In Assumption \ref{asu.FOSD.Y0}, it is stated that the mean potential outcomes of Type 1 and Type 2 surpass those of Type 4 and Type 12 when $D = 0$. Applied to the gender gap, this suggests that the average hourly wages of women in Type 1 and Type 2 exceed those in Type 4 and Type 12. Recall that women in Type 4 and Type 12 do not work, while those in Type 1 always work, and those in Type 2 work if there are no young children at home. The idea behind the assumption is that individuals of Type 1 and Type 2 may exhibit capability traits or are more motivated perhaps. Thus, their performance in the labour market is `better' on average. The assumption only affects the bounds for Type 4 and Type 12, leaving the bounds for Type 1 and Type 2 unchanged. It implies that individuals who work tend to fare better in the labour market, suggesting a form of positive selection. We provide further discussion in Section 4.

In \cite{Zhang2008} and \cite{Blanco2013}, the stochastic dominance assumption is also assumed between types, the always working individuals and the other types. \cite{Zhang2008} assume the distribution of income for the always employed individuals first-order stochastic dominate that for the compliers. \cite{Blanco2013} compare the mean potential outcomes across the two types:  always observed type and compliers.

Assumption \ref{asu.FOSD.Y1} extends the logic to the average hourly wages of men in the gender gap. Similarly, it implies that individuals in Type 1 and Type 2 typically achieve higher mean potential outcomes compared to those in Type 4 and Type 12. However, it suggests that for men, there is a ranking of average hourly wages for different types when individuals are engaged in the labour market.

Both of the assumptions are not directly testable; however, we are able to provide some insights into the mean dominance assumption. Specifically, we identify the mean potential outcomes when $D = 0$ for Type 1 and Type 2. Comparing the estimates for $E(Y^*(0)|\text{Type 1})$ and $E(Y^*(0)|\text{Type 2})$, we have certain intuition about whether women who are always observed have higher potential outcome values than those who work intermittently due to childcare responsibilities. It is also important to note that these insights are based solely on the fundamental assumptions outlined in Assumptions \ref{asu.indep}-\ref{asu.monoZ}. With this comparison, we extend this intuition to other types.

\subsection{Tighter Bounds for Types 1, 2, 4, and 12 under Assumptions \ref{asu.FOSD.Y0} and \ref{asu.FOSD.Y1}}

Assuming the mean dominance assumption for $D = 1$ and $D = 0$ scenarios leads to tighter bounds for all types. Under Assumption \ref{asu.FOSD.Y0}, the bounds of untreated potential outcomes for Types 4 and 12 are narrowed, as $E(Y^*(0)|\text{Type 1})$ and $E(Y^*(0)|\text{Type 2})$ are point identified.\footnote{The bounds under the basic model and Assumption \ref{asu.FOSD.Y0} are presented in the Appendix.} 

Assumption \ref{asu.FOSD.Y1} ensures that in the $(S = 1, Z = 1, D = 1)$ cell, the expected $Y^*(1)$ for Type 1 surpasses the expectation of Types 1, 2, 4, and 12, following the intuition from \cite{Huber2015}. As a result, the lower bound for the potential outcome of Type 1 when treated and $Z = 1$ is $E(Y|D = 1, S = 1, Z = 1)$. Conversely, for Type 12 in the $(S = 1, Z = 1, D = 1)$ cell, the expected $Y^*(1)$ is lower than the expectation for the cell; that is, the upper bound for the potential outcome of Type 12 when treated is $E(Y|D = 1, S = 1, Z = 1)$. 

Similarly, in the $(S = 1, Z = 0, D = 1)$ cell, the expected $Y^*(1)$ for Type 1 outperforms the overall expectation: $E(Y|D = 1, S = 1, Z = 0)$. The expectation for the potential outcome when treated for Type 4 should be lower than the expectation. Whether the expectation for the potential outcome when treated for Type 2 will be lower or higher depends on the proportion of each type.

Overall, under Assumption \ref{asu.FOSD.Y1}, the bounds for the ATE for Type 1 are calculated as follows: $$UB_{T1} = \min_{z} E[Y|D = 1, S = 1, Y \geq Y_{1 - q_{T1_z}}, Z = z]-E(Y^*(0)|\text{Type 1})$$
$$LB_{T1}=  \max_{z} \{E(Y|D = 1, S = 1, Z = z)\}-E(Y^*(0)|\text{Type 1})$$

For Type 2, the lower bounds remain the same. However, the upper bounds when $Z = 1$ and $Z = 0$ change under the first inequality sign of Assumption \ref{asu.FOSD.Y1} since the upper bounds of Type 2 should be lower than the upper bounds of Type 1. The updated bounds for the ATE for Types 2 are as follows: $$UB_{T2} = \min_{z} E[Y|D = 1, S = 1, Y \geq Y_{1 - q_{T1_z}}, Z = z]-E(Y^*(0)|\text{Type 2})$$
$$LB_{T2}=  \max_{z} E[Y|D = 1, S = 1, Y \leq Y_{q_{T2_z}}, Z = z]-E(Y^*(0)|\text{Type 2})$$ 

With Assumptions \ref{asu.FOSD.Y0} and \ref{asu.FOSD.Y1}, for Type 4, the bounds on the ATE are as follows: $$UB_{T4} =  \min \{E(Y|D = 1, S = 1, Z = 0), C\}-Y^{LB}$$
$$LB_{T4}=  \max_{z} E[Y|D = 1, S = 1, Y \leq Y_{q_{T4_z}}, Z = z]- \min_{T \in \{1,2\}}  E(Y^*(0)| \text{Type T})$$

The term $C =E[Y|D = 1, S = 1, Y \geq Y_{1 - q_{T_z}}, Z = z]$. i.e., C is the upper bound of $E(Y^*(1)|\text{Type T})$ for Type 1, Type 2, and Type 4.

For Type 12, the bounds on the ATE are as follows: $$UB_{T12} =  \min_z \{E(Y|D = 1, S = 1, Z = z) \}-Y^{LB}$$
$$LB_{T12}=  E[Y|D = 1, S = 1, Y \leq Y_{q_{T12_0}}, Z = 1]- \min_{T \in \{1,2\}}  E(Y^*(0)| \text{Type T})$$

If we only assume that Type 1's expected potential outcome when treated dominates that of Types 4 and 12, then the upper bounds of Types 4 and 12 are as follows: $$UB_{T} = \min_{z} E[Y|D = 1, S = 1, Y \geq Y_{1 - q_{T1_z}}, Z = z]-Y^{LB}.$$

For estimation, we follow the previous estimation strategy for the trimmed conditional expectations and calculate the sample analogue for the conditional expectations $E(Y|D = 1, S = 1, Z = z)$ when $z \in \{0, 1\}$. The asymptotic distribution also follows Theorem \ref{theoremCAN}.

\section{Empirical Application} \label{sec:emp}

In this section, we apply our method to bound the average treatment effect of gender on the log hourly wage\footnote{The hourly wage is the ratio between the previous year's wage and the hours worked last year. For specific details, see \cite{Maasoumi2019}. The extremely low values of hourly wages are recorded as zeros in the data set. Hence, the missing values of the log of hourly wage variable used in \cite{Maasoumi2019} include those that are missing in the hourly wage and also extremely low values.} for each type based on observed and unobserved characteristics. The data is from \cite{Maasoumi2019} based on CPS in the United States from 1976 to 2013. For each year, the sample size is different. For instance, in 1976, among those 64,249 records, 34,306 are from females and 29943 are from males. In 2013, there are 61,657 female records, and 57,542 are from males. We focus on not only the full-time workers but also the part-time workers, as \cite{Fernandez2023}.

Before we construct our bounds, we need to consider the effects of baseline covariates. In \cite{Maasoumi2019}, those covariates are controls. In our analysis, we follow the method proposed in \cite{Lee2009} to bound the ATE conditional on these observable characteristics. Specifically, we use the same group of covariates, their polynomials, and interaction terms as in \cite{Maasoumi2019} to construct a proxy for predicted hourly wage for each individual by linear regression estimation. This proxy is calculated for each individual, regardless of their employment status. Based on the proxy, we divide the sample of each year into five groups, i.e., covariate groups. After this division, we calculate the proportion and construct the bounds of ATE for each type in every covariate group. Additionally, we consider the weights of each recorded data point, as in \cite{Maasoumi2019} in our calculations for bounds.

In summary, we find that the bounds for Type 1 (EEEE stratum) are narrower than the other types. Also, we see that the LBs and UBs for Type 1 drop rapidly in the early years of the sample (1976–1990s). After that, the bounds sometimes contain 0 and sometimes exclude 0. For Type 2 (EENE stratum), the LBs and UBs of ATE show more and more bounds exclusive to 0. The 95\% Confidence Intervals (CIs) are very wide; most of them contain 0. After we employ the mean dominance assumption, the bounds of ATE for Type 1 exclude the 0 through the period. The UBs decreases display the same trend as under the basic model. For Type 4 (EENN stratum), LBs and the LBs of 95\% CIs are above 0 in the early years for higher predicted wage groups. For Type 12 (ENNN stratum), like Type 2, the bounds of ATE exclude 0 in the later years, yet the 95\% CIs do not.

\subsection{Results under Basic Model}

In this subsection, we present the proportions and the bounds of the ATE for each type under the basic model setting. For the bounds, we show the intersection bounds, CLR bounds, and their corresponding 95\% CIs. Intersection bounds refer to the intersection of the bounds of the ATE under $Z = 0$ and $Z = 1$. However, intersection bounds ignore the covariance between the bounds, leading to bias in finite sample settings. Therefore, we apply the method proposed in \cite{Chernozhukov2013} to obtain a bias-adjusted version, known as CLR bounds (or bounds under CLR adjustment). Specifically, the CLR bounds here are the half-median unbiased estimates of the lower and upper bounds (\cite{Chernozhukov2013}).\footnote{It is noteworthy that Xuan Chen generously provided the STATA code from \cite{ChenXuan2015}, which utilizes the CLR methods from \cite{Chernozhukov2013}. We express our sincere appreciation for this contribution and have generated an R version of the code based on the method outlined in \cite{Chernozhukov2013}.} Correspondingly, there are two types of 95\% CIs. The first one is from \cite{Imbens2004}, calculated as $(\widehat{LB} - 1.645 \widehat{SE_{LB}}, \widehat{UB} + 1.645 \widehat{SE_{UB}})$, where the estimators and their Standard Errors are from Theorem \ref{theoremCAN}. The second one is the CLR 95\% CI, from \cite{Chernozhukov2013}. CLR bounds and CLR 95\% CI will be wider than their counterparts: intersection bounds and 95\% CI from \cite{Imbens2004}. 

\begin{figure}[!htbp]
\caption{Proportions of types}
  \centering
\includegraphics[width=0.9\linewidth]{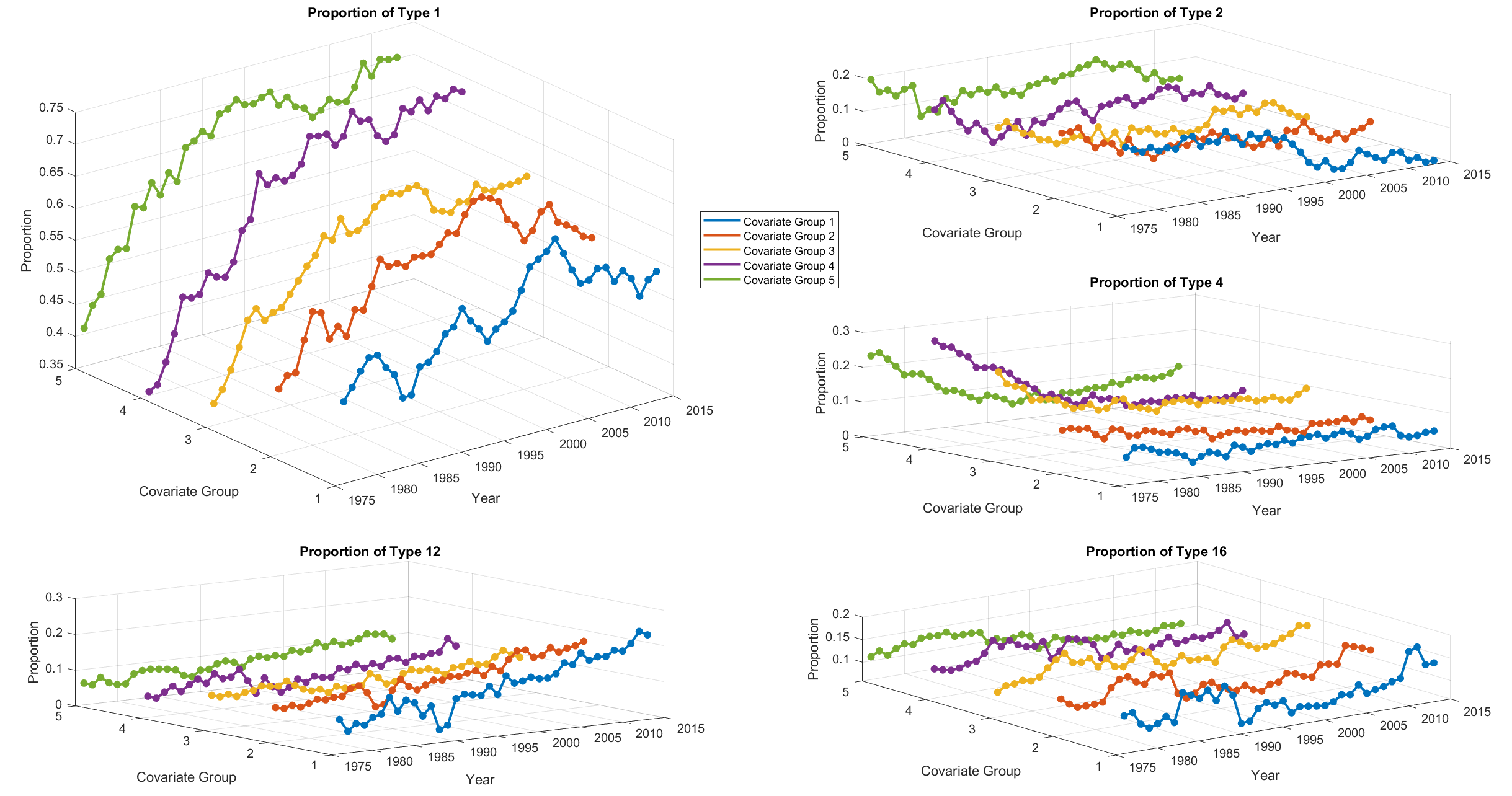} \label{fig:proportions}
\end{figure}

First, we examine the proportions of each type across different years and covariate groups. Figure \ref{fig:proportions} illustrates these proportions for each type, with five separate figures provided for each type. Notably, Type 1 (EEEE stratum) consistently exhibits the highest proportions across all years and covariate groups. This is evident from the axis displaying the proportions, where the lowest bar represents a proportion of 0.35 for Type 1, while the highest possible value for other types is 0.3. The figure for Type 1 also shows that there is a big jump from 1976 to 1990; that is, the proportion of always-working individuals (men and women) rises substantially over those years. The magnitude of this increase is most pronounced in covariate group 5 compared to the other four groups. Additionally, when comparing proportions across covariate groups, we notice that as we ascend towards higher predicted income groups, the proportions of consistently working individuals also increase. This is reasonable, as individuals with higher wages tend to engage in employment. Recall that when we calculate the proportions of Type 1 in the whole population, we use the conditional probability $P_r(S =1| D = 0, Z = 1)$. The rise in $P_r(S =1| D = 0, Z = 1)$ shows that more and more mothers with young children choose to work. This is also noted in \cite{Blau1998}. 

However, for the other four types, the trends vary. For Type 2 (EENE stratum) and Type 4 (EENN stratum), there is a noticeable downward trend across the years for most covariate groups. For Type 2, although it is unclear which covariate group dominates the proportion, we observe a significant decrease in the proportions of Type 2 individuals in covariate group 1 over the years, with proportions nearing zero in later years. Recall that the proportion of Type 2 is calculated as the gap between the proportion of women who work and that of women who choose to work with young children at home, that is, the proportion of women who choose not to work with young children at home. This downward trend indicates a decline in the proportion of women who choose not to work with young children at home. Similarly, for Type 4, there is a substantial decrease in proportions for covariate groups 3, 4, and 5 around the 1990s, followed by a slight increase in covariate group 5.

Type 12 (bottom left figure) exhibits higher proportions in later years, similar to Type 1. However, while the proportion of Type 12 (ENNN stratum) individuals increases over time across all covariate groups, it decreases as we progress from covariate group 1 to covariate group 5. Remember, individuals categorized as Type 12 only work if they are male and have young children at home. This disparity across covariate groups indicates a greater prevalence of Type 12 individuals in the lowest predicted income bracket, which is understandable. Lastly, it is worth noting the steady presence of Type 16 in our model, representing individuals who never work. While we report its proportion, we do not provide the bounds of ATE for Type 16. 

We delve into the bounds of ATEs for Type 1 and Type 2 across five covariate groups for four years: 1976, 1996, 2011, and 2013 under the basic model, reported in Table \ref{table:Type1Type2}, Figure \ref{fig:type1}, and Figure \ref{fig:type1and2} in the Appendix. The choice of 1976 allows us to establish a baseline, while 1996 showcases the pattern in the bounds. For a comprehensive overview of the results spanning 38 years, refer to the tables with complete results in the Supplementary Appendix.

First, let us focus on Type 1. Initially, our attention is drawn to the lower bound (LB) of the intervals for ATEs, as it signifies whether a non-negative gender gap exists. In 1976, the lower bound of CLR on ATE consistently exceeds 0 across all covariate groups, indicating the presence of a gender gap. However, for the first covariate group (the group with predicted lower income levels), the lower bounds are all below 0, implying that 0 is contained within the CIs. Conversely, for all the other four covariate groups, 0 is not included in the CIs. Notably, the trend for Type 1 suggests that the intervals for the gender gap gradually encompass 0 in the later years. Although the trend varies across different covariate groups, the overall direction indicates a gradual inclusion of 0 in the intervals as time progresses. Furthermore, the intervals for the gender gap among individuals in covariate groups 3, 4, and 5 increasingly encompass 0. Notably, covariate groups 3, 4, and 5 represent individuals predicted to have higher income levels. For covariate group 1, the intervals start to include 0 after 1981 and remain inconclusive for most of the sample years. Next, the Upper Bound (UB) for the gender gap in covariate group 1 is consistently the lowest among the five groups. Conversely, covariate group 5 generally exhibits higher UB values. Additionally, UB values decrease gradually over time. This observation aligns with findings from \cite{Maasoumi2019}, \cite{BlauKahn2017}, and \cite{Goldin2014}, indicating that the gender gap has decreased over time, although the decease is not the same across all groups of individuals. For instance, for lower predicted income groups, women do not appear to earn less than men. While in groups with high potential income, the gender gaps are generally larger.

It is essential to note the evolving proportions of individuals in Type 1 over time, ranging from a minimum of 0.356 to a maximum of 0.709. The median proportion stands at 0.596, with the proportion of Type 1 surpassing 0.50 across all five covariate groups after 1985. These relatively high proportions provide us with relatively low standard errors. Thus, there is no big difference between the bounds and CIs. Also, individuals classified as Type 1 always choose to work. In earlier years, these individuals predominantly comprised men who supported the family and women from impoverished backgrounds who married less affluent partners, resulting in substantial gender gaps across the five covariate groups. However, over time, an increasing number of women opt to enter the workforce, leading to a general reduction in the gender gap. 

Figure \ref{fig:type1}, based on the table on CLR bounds for Type 1 in the Appendix, further supports these findings by incorporating gender gap results from \cite{Maasoumi2019} (Table 10 first column: Mean). Notably, the CLR bounds (and the intersection bounds also) contain the mean across five quantiles in Table 10 of \cite{Maasoumi2019} after smoothing by using lowess, suggesting alignment without imposing numerous assumptions.

Next, for Type 2, we see that although 1979 marks the first year when the CLR bounds for high potential wage groups no longer encompass 0, as indicated in the Supplementary Appendix, throughout most of the sample period, the 95\% CIs for the gender gap remain wide and inclusive of 0. This is primarily due to the comparatively low proportion of Type 2 individuals, particularly in later years. This change in the proportion is shown in Figure \ref{fig:proportions}. By examining the numbers, the proportion of Type 2 never exceeds 0.100 after 2008. This low proportion contributes to the wide bounds of the ATE, indicating substantial uncertainty, large standard errors, and wide 95\% CIs.

The CLR bounds and intersection bounds for Type 2 of potential high-wage groups (covariate groups 4 and 5) exclude 0 for more than 20 years, particularly in 1979, 1981, and subsequent years. Additionally, the LB of the gender gap exceeds 0.5 after 2009 for covariate group 5, suggesting a significant gender gap among individuals predicted to have higher income levels. In other words, men probably perform better than women for those covariate groups in those years. Recall that women of Type 2 will choose not to work when they have young children at home and they will work when they do not, and the covariate groups 4 and 5 contain individuals who are predicted to have higher income levels. This suggests that if women of Type 2, who are predicted to have higher wages, work after their children are older than 5, they will face the gender gap issue. For covariate groups 1, 2, and 3, the intervals consistently include 0 across the years, indicating that women in lower predicted income groups likely do not perform worse than men.

The 95\% CIs presented in Table \ref{table:Type1Type2} for Type 2 are wide due to the low proportion, resulting in substantial standard errors. Indeed, with this low proportion, the standard errors of estimates for $E(Y^*(0)|\text{Type 2})$, UB, and LB are large. Consequently, the standard errors of the bounds for ATE are much larger compared to the results for Type 1. As a consequence, the 95\% CIs are wide and typically encompass 0. However, there is an exception in 2010. Upon careful examination of tables in the appendix, it is observed that in 2010, the Imbens and Manski 95\% CI (the tightest CI) provides bounds with a positive LB. Furthermore, comparing the LBs of CIs across the five covariate groups, it is noted that the LBs for covariate groups 4 and 5 are closer to 0 (less negative). Additionally, upon examining across the years, it is observed that the LBs for covariate groups 5 are gradually moving towards 0 from the negative side.

\begin{figure}[!htbp]
\caption{Upper and lower CLR bounds of the Gender Gap for Type 1  across years and 5 covariate groups under basic assumptions}  \label{fig:type1}
  \centering
\includegraphics[width=0.9\linewidth]{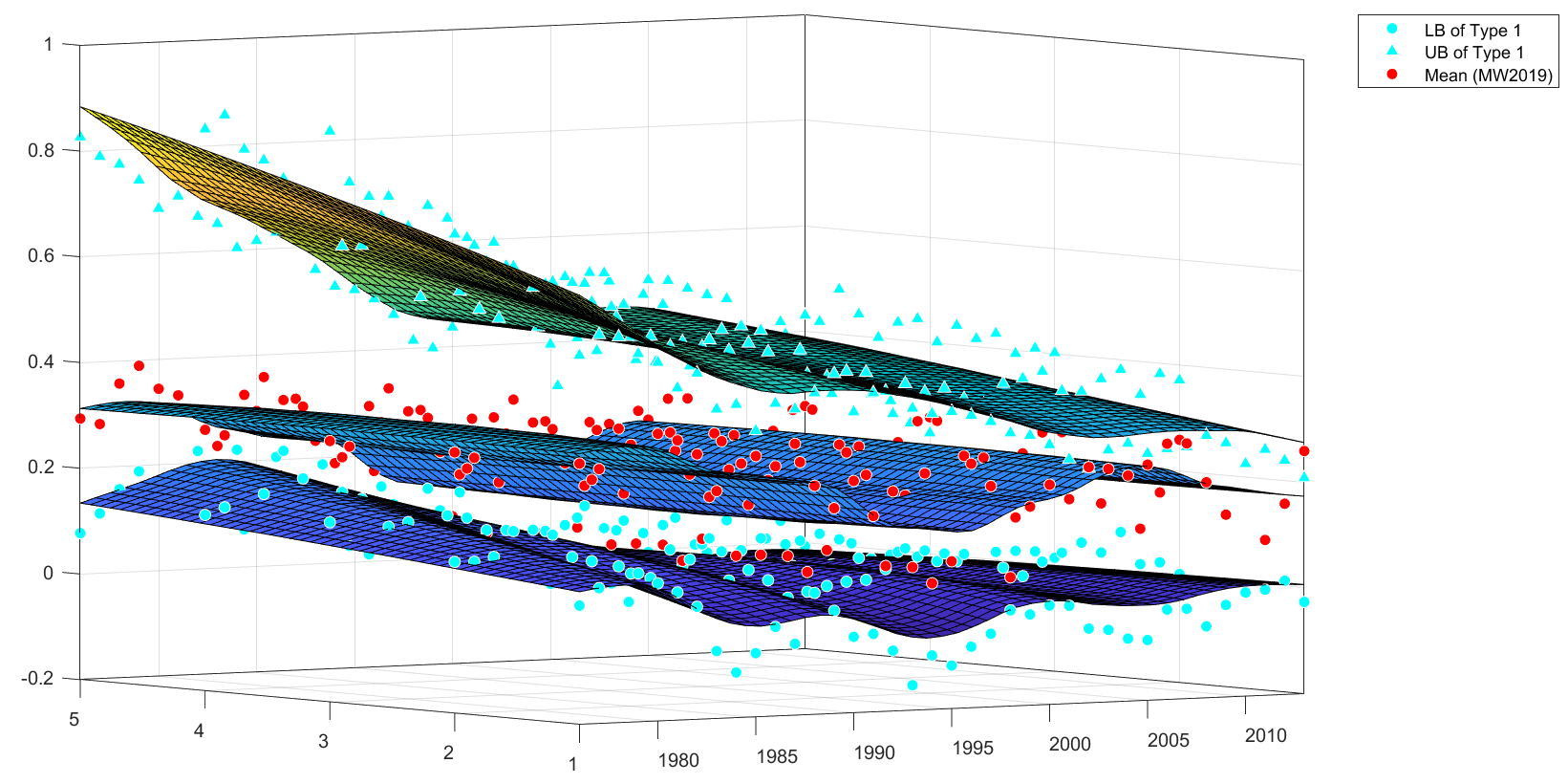}
\end{figure}

Next, we compare Type 1 and Type 2 under the basic model. Figure \ref{fig:type1and2} is drawn to provide a clear depiction of the trend.\footnote{In the tables for Type 2 and covariate group 4, there are 3 big numbers in LB in 1983, 1987, and 2012. This happens because the estimates for $E(Y(0)|Type 2)$ of covariate group 4 in the years 1983, 1987, 2012, and 2013 are below 1. This occurrence is attributed to the very low proportion of Type 2 in the dataset. Across the five covariate groups, such situations only arise in covariate group 4. Therefore, these cases are excluded from our analysis.} This figure presents the bounds of ATEs for Type 1 and Type 2 across years and covariate groups. Based on Table \ref{table:Type1Type2}, it is observed that the bounds for the gender gap for Type 1 are much narrower compared to those for Type 2. This discrepancy is reasonable because the proportion of Type 2 is significantly lower than that of Type 1, indicating fewer individuals belong to Type 2. Consequently, the upper bounds of the gender gap for individuals belonging to Type 2 are progressively higher from 1976 to 2013. However, the lower bounds do not decrease notably, especially for covariate group 5. Moreover, the standard errors for Type 2 are higher than those for Type 1, resulting in wider 95\% CIs for Type 2 compared to Type 1.

\begin{table}[!htbp]
  \centering
   \scalebox{.70}{
\begin{tabular}{cccccc}
\hline
\hline
\multicolumn{6}{l}{\textbf{PANEL A: Type 1 under Basic Assumptions in 1976 across 5 covariate groups}} \\ \hline
Covariate Group & 1 &2 &3 &4 &5 \\ \hline
Bounds	&[ LB,	UB]&[	LB,	UB]&[	LB,	UB]&[	LB,	UB]&[	LB,	UB]\\ \hline
Intersection Bounds  &	[ 0.036, 0.491]&[	0.095, 0.700]&[ 0.148, 0.874]&[	0.141,	0.856]&[	0.085,	0.819]\\
CLR Bounds                  &[ 0.025, 0.499]&[	0.087, 0.706]&[ 0.140, 0.880]&[	0.133,	0.863]&[	0.077,	0.826]\\
Imbens and Manski 95\% CI: &[	-0.015,	0.529]&[	0.055,	0.733]&[	0.104,	0.915]&[	0.088,	0.909]&[	0.028,	0.877]\\
 CLR 95\% CI	&[	-0.024,	0.536]&[	0.049,	0.738]&[	0.097,	0.920]&[	0.081,	0.916]&[	0.021,	0.884]\\ \hline
 \multicolumn{6}{l}{\textbf{PANEL A: Type 2 under Basic Assumptions in 1976 across 5 covariate groups}} \\ \hline
Intersection Bounds  &	[ -0.291, 0.766]&[ -0.430, 0.734]&[ -0.387, 0.858]&[ -0.048, 1.143]&[ -0.166, 1.097]\\
CLR Bounds                  &[ -0.304, 0.778]&[ -0.441, 0.745]&[ -0.399, 0.869]&[ -0.062, 1.152]&[ -0.177, 1.107]\\
Imbens and Manski 95\% CI: &[ -0.389, 0.857]&[ -0.525, 0.821]&[ -0.518, 0.983]&[ -0.179, 1.267]&[ -0.295, 1.223]\\
 CLR 95\% CI	&[ -0.403, 0.870]&[ -0.537, 0.832]&[ -0.534, 0.998]&[ -0.196, 1.281]&[ -0.309, 1.236]\\ \hline
 \multicolumn{6}{l}{\textbf{PANEL B: Type 1 under Basic Assumptions in 1996 across 5 covariate groups}} \\ \hline
Intersection Bounds  &[ -0.076, 0.346]&[ 0.024, 0.408]&[ 0.061, 0.465]&[ 0.033, 0.404]&[ 0.007, 0.459]\\
CLR Bounds &[ -0.084, 0.353]&[ 0.018, 0.415]&[ 0.054, 0.473]&[ 0.025, 0.412]&[ 0.000, 0.467]\\
Imbens and Manski 95\% CI: &[ -0.117, 0.381]&[ -0.012, 0.445]&[ 0.017, 0.509]&[ -0.016, 0.448]&[ -0.040, 0.505]\\
CLR 95\% CI	&[ -0.124, 0.388]&[ -0.018, 0.451]&[ 0.010, 0.517]&[ -0.025, 0.457]&[ -0.048, 0.512]\\ \hline
\multicolumn{6}{l}{\textbf{PANEL B: Type 2 under Basic Assumptions in 1996 across 5 covariate groups}} \\ \hline
Intersection Bounds  &[ -0.574, 0.818]&[ -0.823, 0.859]&[ -0.199, 2.090]&[ 0.144, 2.018]&[ 0.190, 2.382]\\
CLR Bounds &[ -0.584, 0.834]&[ -0.835, 0.880]&[ -0.212, 2.136]&[ 0.129, 2.045]&[ 0.173, 2.406]\\
Imbens and Manski 95\% CI: &[ -0.715, 0.966]&[ -1.043, 1.087]&[ -0.774, 2.696]&[ -0.304, 2.483]&[ -0.212, 2.792]\\
CLR 95\% CI	&[ -0.731, 0.985]&[ -1.068, 1.119]&[ -0.855, 2.807]&[ -0.369, 2.560]&[ -0.265, 2.851]\\ \hline
 \multicolumn{6}{l}{\textbf{PANEL C: Type 1 under Basic Assumptions in 2011 across 5 covariate groups}} \\ \hline
Intersection Bounds  &[ 0.008, 0.265]&[ 0.033, 0.341]&[ 0.037, 0.398]&[ 0.009, 0.287]&[ 0.017, 0.381]\\
CLR Bounds &[ 0.000, 0.265]&[ 0.027, 0.348]&[ 0.031, 0.404]&[ 0.002, 0.293]&[ 0.011, 0.387]\\
Imbens and Manski 95\% CI: &[ -0.029, 0.302]&[ 0.001, 0.374]&[ -0.002, 0.438]&[ -0.025, 0.321]&[ -0.016, 0.415]\\
CLR 95\% CI	&[ -0.037, 0.304]&[ -0.005, 0.380]&[ -0.009, 0.445]&[ -0.032, 0.328]&[ -0.022, 0.421]\\ \hline
\multicolumn{6}{l}{\textbf{PANEL C: Type 2 under Basic Assumptions in 2011 across 5 covariate groups}} \\ \hline
Intersection Bounds  &[ -1.382, 0.934]&[ -0.959, 1.097]&[ -0.525, 1.533]&[ 0.085, 2.337]&[ 0.803, 3.550]\\
CLR Bounds &[ -1.393, 0.989]&[ -0.970, 1.122]&[ -0.538, 1.553]&[ 0.068, 2.360]&[ 0.787, 3.573]\\
Imbens and Manski 95\% CI: &[ -2.125, 1.742]&[ -1.258, 1.413]&[ -0.928, 1.943]&[ -0.514, 2.936]&[ -0.213, 4.569]\\
CLR 95\% CI	&[ -2.238, 1.902]&[ -1.292, 1.460]&[ -0.980, 2.003]&[ -0.603, 3.030]&[ -0.375, 4.737]\\ \hline
\multicolumn{6}{l}{\textbf{PANEL D: Type 1 under Basic Assumptions in 2013 across 5 covariate groups}} \\ \hline
Intersection Bounds  &[ -0.019, 0.208]&[ 0.011, 0.365]&[ 0.021, 0.395]&[ -0.067, 0.223]&[ 0.000, 0.424]\\
CLR Bounds &[ -0.026, 0.208]&[ 0.005, 0.372]&[ 0.015, 0.402]&[ -0.075, 0.229]&[ -0.006, 0.431]\\
Imbens and Manski 95\% CI: &[ -0.056, 0.246]&[ -0.021, 0.398]&[ -0.021, 0.438]&[ -0.108, 0.259]&[ -0.036, 0.461]\\
CLR 95\% CI	&[ -0.065, 0.249]&[ -0.027, 0.404]&[ -0.029, 0.446]&[ -0.116, 0.266]&[ -0.042, 0.467]\\  \hline
\multicolumn{6}{l}{\textbf{PANEL D: Type 2 under Basic Assumptions in 2013 across 5 covariate groups}} \\ \hline
Intersection Bounds  &[ -1.568, 1.313]&[ -0.962, 0.943]&[ -0.149, 2.321]&[ 0.807, 3.143]&[ 0.873, 3.795]\\
CLR Bounds &[ -1.583, 1.445]&[ -0.972, 0.961]&[ -0.165, 2.362]&[ 0.790, 3.170]&[ 0.854, 3.824] \\
Imbens and Manski 95\% CI: &[ -3.691, 3.701]&[ -1.184, 1.176]&[ -1.118, 3.327]&[ -0.074, 4.026]&[ -0.255, 4.936]\\
CLR 95\% CI	&[ -4.073, 4.229]&[ -1.206, 1.206]&[ -1.276, 3.511]&[ -0.217, 4.177]&[ -0.437, 5.130]\\
\hline
\hline
\end{tabular}
}
\caption{Bounds on ATE comparison for Type 1 and Type 2}\label{table:Type1Type2}
\end{table}

The discussion concerning the ATE for Types 4 and 12 is deferred to a later section. Under the basic model, we rely on theoretical bounds for $E(Y^*(0)| \text{Type j})$ when $j \in \{4, 12\}$. Although the bounds for the ATE for Types 4 and 12 will be considerably wide, they provide valuable information as they are narrower than $[Y^{LB} - Y^{UB}, Y^{UB} - Y^{LB}]$.

\subsection{Check the Mean Dominance Assumption Using the Data under Basic Model}

In this subsection, we compare the estimates of $E(Y^*(0)| \text{Type j})$ when $j \in \{1, 2\}$ to provide a check for the mean dominance assumption for those types. The mean dominance assumption is related to the positive selection assumption. However, the mean dominance assumption compares wages based on types, and positive selection relies on working statuses. For instance, women of Type 2 (EENE stratum) have two possible working statuses: employed and nonemployed. $E(Y^*(0)| \text{Type 1}) > E(Y^*(0)| \text{Type 2})$ implies that wages for Type 1 (EEEE stratum) tend to be higher than those for Type 2 (EENE stratum).

Positive selection, explored in the literature, posits that wages for employed women tend to be higher than those for non-working women. This assumption has been discussed in works such as \cite{BlauKahn2006}, \cite{Blundell2007}, \cite{Olivetti2008}, \cite{Mulligan2008}, \cite{BlauKahn2017}, and \cite{Maasoumi2019}. \cite{Blau2023} look into the pattern of selection into the employed status on the gender gap either in the median, mean, or the whole distribution. Their conclusions vary based on methods, including part-time workers or not, the definition of selection, sample data sets, and models. For instance, \cite{Mulligan2008} is based on the model and method from \cite{Heckman1979}, and \cite{Maasoumi2019} relies on nonseparate models with the copula method. \cite{Blundell2007} assume that for men and women, the wage distribution of nonworkers is first-order stochastically dominated by the distribution of workers. They also provide a weaker version by comparing the median wage of workers and nonworkers. \cite{Maasoumi2019} test this positive selection from \cite{Blundell2007} using the CPS in the United States. They find that the CDF of the log hourly wage of working women does not always first order stochastic dominate the CDF of not working women. They conclude that for women, the selection changes from negative selection to positive selection. The turning time is the 1990s. For men, \cite{Maasoumi2019} conclude that positive selection through the years.

\begin{figure}[!htbp]
\caption{Untreated Potential Outcome Comparison}
  \centering
\includegraphics[width=0.9\linewidth]{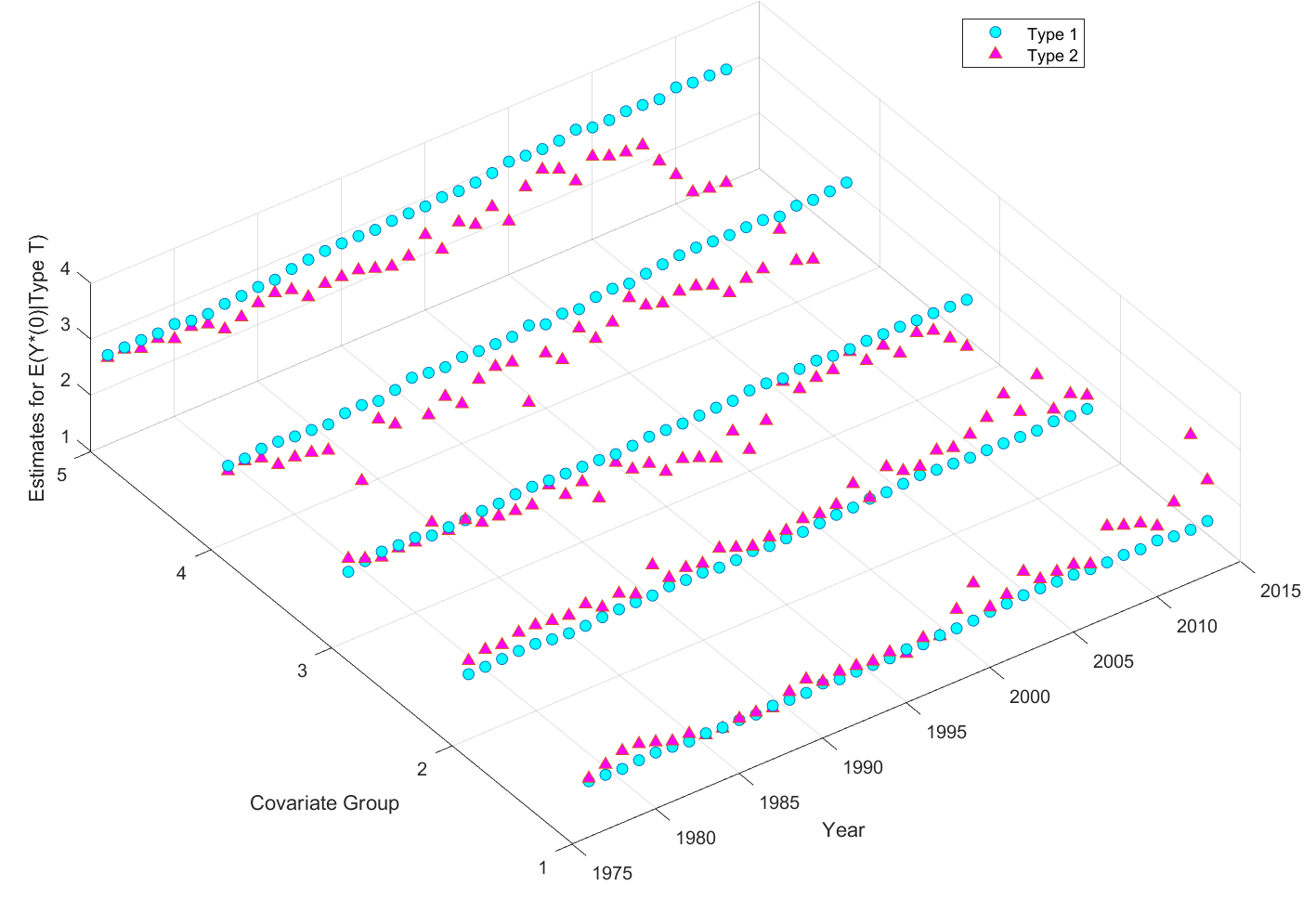} \label{fig:y0Type1and2}
\end{figure}

Figure \ref{fig:y0Type1and2} illustrates the average potential outcomes for women of Type 1 and Type 2 across five covariate groups and years. From the figure, it is evident that for covariate groups 4 and 5, estimates for $E(Y^*(0)| \text{Type 1})$ surpass those of women of Type 2 over the years. Conversely, for covariate groups 1 and 2, the trend is reversed, indicating a notable distinction between Type 1 and Type 2. In covariate group 3, the mean potential outcome of women in Type 1 is initially lower than that of Type 2 for the first two years, but the trend reverses thereafter. 

When considering standard errors and 95\% confidence intervals, the intervals of the estimates overlap in the early years for covariate groups 4 and 5, while for low potential wage groups, the confidence intervals do not always intersect. Consequently, if we calculate the average across the five covariate groups, the average potential outcomes for women of Type 2 are expected to be higher than those of Type 1 in the early years, as indicated by non-overlapping confidence intervals during the first two years. Regarding selection patterns, negative selection is initially observed, followed by positive selection, as depicted in Table \ref{table:T1T2y0} in Appendix. This is a similar conclusion as in \cite{Maasoumi2019} and \cite{Mulligan2008}. \cite{Fernandez2023} analyze the role of selection for CPS data from 1975 to 2020 for all workers using the semiparametric method to estimate functions of distributions based on nonseparate models. They find that there is positive selection, and the selection evolves through time. We consider individuals with full-time jobs and part-time jobs, as in \cite{Fernandez2023}, and find that the pattern changes in 1979, earlier than \cite{Maasoumi2019} observe. Based on the information from Type 1 and Type 2, positive selection predominantly characterizes the pattern, particularly for high-potential outcome groups and later years.

\subsection{Results under Basic Model with Assumptions \ref{asu.FOSD.Y0} and \ref{asu.FOSD.Y1}}

In this subsection, we present the bounds of ATE for all four types under Assumptions \ref{asu.indep}-\ref{asu.FOSD.Y1}. With these additional assumptions, we have narrower bounds and 95\% CIs compared to those under the basic model alone. We provide results for selected years, with complete outcomes available in the Supplementary Appendix.

Table \ref{table:type1type2comSD} reports the results for Type 1 and Type 2 across 5 covariate groups under the basic model and Assumption \ref{asu.FOSD.Y1}. Recall that for Types 1 and 2, $E(Y^*(0)| \text{Type j})$ is point identified. Hence, the bounds of ATE for Types 1 and 2 under the basic model and Assumption \ref{asu.FOSD.Y0} are the same as those only under the basic model. With Assumption \ref{asu.FOSD.Y1}, the bounds for these two types are narrower than before. We still select the same years to make a comparison with Table \ref{table:Type1Type2}.

First, let us focus on Type 1 in Table \ref{table:type1type2comSD}. We see that the LBs for Type 1 are higher with Assumption \ref{asu.FOSD.Y1}. That is, every interval is above 0. The LB and UB are decreasing fast from 1976 to the 1990s. After that, it becomes slower. In some years, the LB and UB even increased. Although there are some differences across the covariate groups, the overall trend stays the same. Among those covariate groups, the highest LBs and UBs happen mostly in covariate group 5. This shows that in the potential high-income group, women who always work probably have relatively poor performance in the labour market compared with men.

Next, Table \ref{table:type1type2comSD} shows that for Type 2, men are probably performing better than women in the later years of the sample for people who are predicted to have higher income. The UBs are lower than those under the basic model. However, UB fluctuates. This may be due to the low proportion of individuals who belong to Type 2. For covariate groups 1 and 2, we observed negative UBs for some years. This shows that women probably perform better than men in the group where people are predicted to have lower income levels. However, if we check the 95\% confidence intervals we see that both those with and those without adjustment using the CLR method contain 0.
\begin{table}[!htbp]
  \centering
   \scalebox{.70}{
\begin{tabular}{cccccc}
\hline
\hline
\multicolumn{6}{l}{\textbf{PANEL A: Type 1 in 1976 across 5 covariate groups}} \\ \hline
Covariate Group & 1 &2 &3 &4 &5 \\ \hline
Bounds	&[ LB,	UB]&[	LB,	UB]&[	LB,	UB]&[	LB,	UB]&[	LB,	UB]\\ \hline
Intersection Bounds  &[ 0.344, 0.491]&[ 0.431, 0.700]&[ 0.517, 0.874]&[ 0.516, 0.856]&[ 0.479, 0.819]\\
CLR Bounds                  &[ 0.335, 0.499]&[ 0.424, 0.706]&[ 0.511, 0.880]&[ 0.510, 0.863]&[ 0.474, 0.826]\\
Imbens and Manski 95\% CI: &[ 0.302, 0.529]&[ 0.398, 0.733]&[ 0.476, 0.915]&[ 0.471, 0.909]&[ 0.431, 0.877]\\
 CLR 95\% CI	&[ 0.290, 0.539]&[ 0.391, 0.739]&[ 0.470, 0.922]&[ 0.464, 0.919]&[ 0.422, 0.887]\\ \hline
\multicolumn{6}{l}{\textbf{PANEL A: Type 2 in 1976 across 5 covariate groups}} \\ \hline
Intersection Bounds  &[ -0.291, 0.439]&[ -0.430, 0.460]&[ -0.387, 0.641]&[ -0.048, 0.951]&[ -0.166, 0.869]\\
CLR Bounds                  &[ -0.304, 0.448]&[ -0.441, 0.468]&[ -0.399, 0.648]&[ -0.062, 0.958]&[ -0.177, 0.876]\\
Imbens and Manski 95\% CI: &[ -0.389, 0.522]&[ -0.525, 0.543]&[ -0.518, 0.757]&[ -0.179, 1.069]&[ -0.295, 0.987]\\
 CLR 95\% CI	&[ -0.406, 0.534]&[ -0.538, 0.553]&[ -0.536, 0.769]&[ -0.198, 1.081]&[ -0.311, 1.000]\\ \hline
 \multicolumn{6}{l}{\textbf{PANEL B: Type 1 in 1996 across 5 covariate groups}} \\ \hline
 Intersection Bounds  &[ 0.213, 0.346]&[ 0.235, 0.408]&[ 0.270, 0.465]&[ 0.242, 0.404]&[ 0.267, 0.459]\\
CLR Bounds &[ 0.205, 0.353]&[ 0.228, 0.415]&[ 0.262, 0.473]&[ 0.235, 0.412]&[ 0.259, 0.467]\\
Imbens and Manski 95\% CI: &[ 0.176, 0.381]&[ 0.200, 0.445]&[ 0.225, 0.509]&[ 0.197, 0.448]&[ 0.221, 0.505]\\
CLR 95\% CI	&[ 0.166, 0.391]&[ 0.192, 0.454]&[ 0.217, 0.519]&[ 0.186, 0.459]&[ 0.211, 0.515]\\ \hline
\multicolumn{6}{l}{\textbf{PANEL B: Type 2 in 1996 across 5 covariate groups}} \\ \hline
Intersection Bounds  &[ -0.574, 0.233]&[ -0.823, 0.179]&[ -0.199, 1.095]&[ 0.144, 1.230]&[ 0.190, 1.396]\\
CLR Bounds &[ -0.584, 0.242]&[ -0.835, 0.186]&[ -0.212, 1.102]&[ 0.129, 1.237]&[ 0.173, 1.404]\\
Imbens and Manski 95\% CI: &[ -0.715, 0.366]&[ -1.043, 0.388]&[ -0.774, 1.660]&[ -0.304, 1.669]&[ -0.212, 1.783]\\
CLR 95\% CI	&[ -0.737, 0.386]&[ -1.077, 0.417]&[ -0.873, 1.752]&[ -0.383, 1.740]&[ -0.281, 1.841]\\ \hline
\multicolumn{6}{l}{\textbf{PANEL C: Type 1 in 2011 across 5 covariate groups}} \\ \hline
Intersection Bounds  &[ 0.246, 0.265]&[ 0.221, 0.341]&[ 0.217, 0.398]&[ 0.185, 0.287]&[ 0.201, 0.381]\\
CLR Bounds &[ 0.236, 0.265]&[ 0.215, 0.348]&[ 0.212, 0.404]&[ 0.179, 0.293]&[ 0.195, 0.387]\\
Imbens and Manski 95\% CI: &[ 0.204, 0.302]&[ 0.190, 0.374]&[ 0.185, 0.438]&[ 0.150, 0.321]&[ 0.173, 0.415]\\
CLR 95\% CI	&[ 0.191, 0.308]&[ 0.182, 0.383]&[ 0.178, 0.447]&[ 0.140, 0.331]&[ 0.167, 0.423]\\ \hline
\multicolumn{6}{l}{\textbf{PANEL C: Type 2 in 2011 across 5 covariate groups}} \\ \hline
Intersection Bounds  &[ -1.382, -0.343]&[ -0.959, 0.129]&[ -0.525, 0.709]&[ 0.085, 1.349]&[ 0.803, 2.337]\\
CLR Bounds &[ -1.393, -0.335]&[ -0.970, 0.135]&[ -0.538, 0.715]&[ 0.068, 1.354]&[ 0.787, 2.341]\\
Imbens and Manski 95\% CI: &[ -2.125, 0.396]&[ -1.258, 0.419]&[ -0.928, 1.099]&[ -0.514, 1.934]&[ -0.213, 3.344]\\
CLR 95\% CI	&[ -2.262, 0.530]&[ -1.305, 0.461]&[ -0.994, 1.157]&[ -0.622, 2.030]&[ -0.400, 3.516]\\ \hline
\multicolumn{6}{l}{\textbf{PANEL D: Type 1 in 2013 across 5 covariate groups}} \\ \hline
Intersection Bounds  &[ 0.177, 0.208]&[ 0.220, 0.365]&[ 0.208, 0.395]&[ 0.150, 0.223]&[ 0.216, 0.424]\\
CLR Bounds &[ 0.168, 0.208]&[ 0.214, 0.372]&[ 0.202, 0.402]&[ 0.143, 0.229]&[ 0.209, 0.431]\\
Imbens and Manski 95\% CI: &[ 0.135, 0.246]&[ 0.189, 0.398]&[ 0.173, 0.438]&[ 0.113, 0.259]&[ 0.179, 0.461]\\
CLR 95\% CI	&[ 0.122, 0.253]&[ 0.181, 0.406]&[ 0.165, 0.448]&[ 0.102, 0.269]&[ 0.177, 0.469]\\ \hline
\multicolumn{6}{l}{\textbf{PANEL D: Type 2 in 2013 across 5 covariate groups}} \\ \hline
Intersection Bounds  &[ -1.568, -0.522]&[ -0.962, 0.125]&[ -0.149, 1.234]&[ 0.807, 2.083]&[ 0.873, 2.450]\\
CLR Bounds &[ -1.583, -0.518]&[ -0.972, 0.132]&[ -0.165, 1.239]&[ 0.790, 2.087]&[ 0.854, 2.454]\\
Imbens and Manski 95\% CI: &[ -3.691, 1.592]&[ -1.184, 0.339]&[ -1.118, 2.191]&[ -0.074, 2.954]&[ -0.255, 3.565]\\
CLR 95\% CI	&[ -4.106, 1.996]&[ -1.216, 0.366]&[ -1.297, 2.357]&[ -0.239, 3.105]&[ -0.464, 3.758]\\
\hline
\hline
\end{tabular}
}
\caption{Bounds under Assumptions \ref{asu.indep}-\ref{asu.monoZ}, and \ref{asu.FOSD.Y1}}\label{table:type1type2comSD}
\end{table}

\begin{table}[!htbp]
  \centering
   \scalebox{.70}{
\begin{tabular}{cccccc}
\hline
\hline
\multicolumn{6}{l}{\textbf{PANEL A: Type 4 in 1976 across 5 covariate groups}} \\ \hline
Covariate Group & 1 &2 &3 &4 &5 \\ \hline
Bounds	&[ LB,	UB]&[	LB,	UB]&[	LB,	UB]&[	LB,	UB]&[	LB,	UB]\\ \hline
Intersection Bounds  &[ -0.474, 1.188]&[ -0.325, 1.534]&[ 0.003, 1.743]&[ 0.185, 1.887]&[ -0.095, 2.074]\\
CLR Bounds &[ -0.508, 1.188]&[ -0.351, 1.534]&[ -0.023, 1.743]&[ 0.134, 1.887]&[ -0.147, 2.074]\\
Imbens and Manski 95\% CI: &[ -0.558, 1.202]&[ -0.389, 1.549]&[ -0.051, 1.755]&[ 0.069, 1.898]&[ -0.213, 2.087]\\
 CLR 95\% CI	&[ -0.580, 1.202]&[ -0.406, 1.549]&[ -0.062, 1.755]&[ 0.043, 1.898]&[ -0.215, 2.087]\\ \hline
 \multicolumn{6}{l}{\textbf{PANEL A: Type 12 in 1976 across 5 covariate groups}} \\ \hline
Intersection Bounds  &[ -0.431, 1.188]&[ -0.394, 1.534]&[ -0.294, 1.743]&[ -0.376, 1.859]&[ -0.528, 2.028]\\
CLR Bounds  & [ -0.452, 1.188]&[ -0.413, 1.539]&[ -0.317, 1.747]&[ -0.415, 1.866]&[ -0.569, 2.035]\\
Imbens and Manski 95\% CI: &[ -0.522, 1.202]&[ -0.473, 1.549]&[ -0.371, 1.755]&[ -0.553, 1.881]&[ -0.691, 2.050]\\
 CLR 95\% CI	&[ -0.537, 1.202]&[ -0.487, 1.552]&[ -0.386, 1.758]&[ -0.583, 1.886]&[ -0.720, 2.054]\\ \hline
 \multicolumn{6}{l}{\textbf{PANEL B: Type 4 in 1996 across 5 covariate groups}} \\ \hline
 Intersection Bounds  &[ -0.606, 1.050]&[ -0.691, 1.393]&[ -0.006, 1.618]&[ 0.151, 1.774]&[ 0.082, 2.124]\\
CLR Bounds &[ -0.636, 1.050]&[ -0.723, 1.393]&[ -0.229, 1.618]&[ -0.017, 1.774]&[ 0.063, 2.124]\\
Imbens and Manski 95\% CI: &[ -0.677, 1.063]&[ -0.764, 1.408]&[ -0.573, 1.632]&[ -0.292, 1.788]&[ -0.313, 2.139]\\
CLR 95\% CI	&[ -0.696, 1.063]&[ -0.784, 1.408]&[ -0.705, 1.634]&[ -0.415, 1.789]&[ -0.364, 2.141]\\ \hline
 \multicolumn{6}{l}{\textbf{PANEL B: Type 12 in 1996 across 5 covariate groups}} \\ \hline
 Intersection Bounds  &[ -0.410, 1.050]&[ -0.538, 1.393]&[ -0.081, 1.618]&[ 0.076, 1.774]&[ 0.101, 2.124]\\
CLR Bounds &[ -0.427, 1.054]&[ -0.558, 1.398]&[ -0.277, 1.622]&[ -0.077, 1.779]&[ 0.101, 2.129]\\
Imbens and Manski 95\% CI: &[ -0.460, 1.063]&[ -0.595, 1.408]&[ -0.654, 1.632]&[ -0.369, 1.788]&[ -0.295, 2.139]\\
CLR 95\% CI	&[ -0.470, 1.066]&[ -0.606, 1.411]&[ -0.816, 1.636]&[ -0.489, 1.791]&[ -0.331, 2.144]\\ \hline
\multicolumn{6}{l}{\textbf{PANEL C: Type 4 in 2011 across 5 covariate groups}} \\ \hline
Intersection Bounds  &[ -0.706, 1.107]&[ -0.758, 1.448]&[ -0.432, 1.652]&[ 0.081, 1.850]&[ 0.997, 2.245]\\
CLR Bounds &[ -0.728, 1.107]&[ -0.788, 1.448]&[ -0.587, 1.652]&[ -0.128, 1.850]&[ 0.981, 2.245]\\
Imbens and Manski 95\% CI: &[ -0.754, 1.118]&[ -0.826, 1.460]&[ -0.823, 1.663]&[ -0.511, 1.861]&[ -0.012, 2.258]\\
CLR 95\% CI	&[ -0.767, 1.118]&[ -0.844, 1.460]&[ -0.810, 1.663]&[ -0.686, 1.863]&[ -0.194, 2.260]\\ \hline
\multicolumn{6}{l}{\textbf{PANEL C: Type 12 in 2011 across 5 covariate groups}} \\ \hline
Intersection Bounds  &[ -0.379, 1.107]&[ -0.472, 1.448]&[ -0.315, 1.631]&[ 0.248, 1.850]&[ 1.109, 2.240]\\
CLR Bounds &[ -0.398, 1.111]&[ -0.488, 1.452]&[ -0.460, 1.638]&[ 0.047, 1.854]&[ 1.109, 2.247]\\
Imbens and Manski 95\% CI: &[ -0.433, 1.118]&[ -0.515, 1.460]&[ -0.707, 1.654]&[ -0.342, 1.861]&[ 0.101, 2.262]\\
CLR 95\% CI	&[ -0.443, 1.121]&[ -0.523, 1.462]&[ -0.702, 1.659]&[ -0.513, 1.865]&[ -0.071, 2.262]\\ \hline
\multicolumn{6}{l}{\textbf{PANEL D: Type 4 in 2013 across 5 covariate groups}} \\ \hline
Intersection Bounds  &[ -0.679, 1.085]&[ -0.795, 1.416]&[ 0.150, 1.629]&[ 0.880, 1.827]&[ 1.168, 2.241]\\
CLR Bounds &[ -0.710, 1.085]&[ -0.823, 1.416]&[ -0.220, 1.629]&[ 0.862, 1.827]&[ 1.153, 2.241]\\
Imbens and Manski 95\% CI: &[ -0.745, 1.096]&[ -0.858, 1.428]&[ -0.807, 1.640]&[ 0.005, 1.838]&[ 0.052, 2.254]\\
CLR 95\% CI	&[ -0.763, 1.096]&[ -0.875, 1.428]&[ -0.751, 1.641]&[ -0.159, 1.840]&[ -0.153, 2.256]\\ \hline
\multicolumn{6}{l}{\textbf{PANEL D: Type 12 in 2013 across 5 covariate groups}} \\ \hline
Intersection Bounds  &[ -0.369, 1.085]&[ -0.489, 1.416]&[ 0.138, 1.629]&[ 1.021, 1.827]&[ 1.122, 2.241]\\
CLR Bounds &[ -0.388, 1.089]&[ -0.505, 1.420]&[ -0.207, 1.632]&[ 1.021, 1.831]&[ 1.122, 2.245]\\
Imbens and Manski 95\% CI: &[ -0.420, 1.096]&[ -0.534, 1.428]&[ -0.821, 1.654]&[ 0.148, 1.838]&[ 0.005, 2.254]\\
CLR 95\% CI	&[ -0.429, 1.098]&[ -0.543, 1.430]&[ -0.795, 1.643]&[ -0.006, 1.842]&[ -0.190, 2.258]\\
\hline
\hline
\end{tabular}
}
\caption{Bounds on ATE comparison for Types 4 and 12 under Assumptions \ref{asu.indep}-\ref{asu.FOSD.Y1}}\label{table:type4type12comSD}
\end{table}

Table \ref{table:type4type12comSD} reports the bounds for Type 4 and Type 12 under Assumptions \ref{asu.indep}-\ref{asu.FOSD.Y1}. That is, to have the tightest bounds for Type 4 and Type 12, we need both \ref{asu.FOSD.Y0} and \ref{asu.FOSD.Y1}. This is because Assumption \ref{asu.FOSD.Y0} provides the upper bounds of $E(Y^*(0)| \text{Type j})$ when $j \in \{4, 12\}$. Assumption \ref{asu.FOSD.Y1}, on the other hand, helps shrink the bounds for $E(Y^*(1)| \text{Type j})$ when $j \in \{4, 12\}$. In the Supplementary Appendix, we report the additional results for Type 4 and Type 12 under the basic model with only Assumption \ref{asu.FOSD.Y0}. The bounds are wider than the bounds we report and discuss in this subsection. Specifically, the UBs tend to be higher. Just as in Table \ref{table:type1type2comSD}, we only report the results for several years to show the pattern.

Table \ref{table:type4type12comSD} presents the same measures and categories for Type 4 across 5 covariate groups and 4 selected years. In 1976, CLR bounds and CLR 95\% CI do not contain 0 when we look at the higher potential income group: covariate group 4. In fact, in the beginning of the sample period (1976--1981), it is quite clear to see that the CLR bound and CLR 95\% CI of ATE do not contain 0 for individuals in covariate group 4 if we check the tables in the Supplementary Appendix. That is, we see that in 1976--1981, the gender gap existed in the predicted high-income group (covariate group 4), and the LBs were closer to 0 from the left-hand side (covariate group 5). Recall that women of Type 4 will not work even if they do not have young children at home, and men, on the other hand, will work regardless of whether they have young children or not. In the early years, well-educated women or potential high-wage women may choose to stay at home after they are married. As noted in \cite{Neal2004}, a high proportion of women in earlier years married to men with high wages and chose to stay at home. Even if they choose to work at that time, their performance in the labour market will probably not be good compared to men. This clear message benefits from the fact that the proportion of Type 4 in that covariate group was the highest at the beginning of the sample period. This is shown in Figure \ref{fig:proportions}. From 1976 to 1995, 0 was not in the CLR bounds for the covariate group 4 for 12 years (excluding 1983 and 1987). From 1996 to 2008, CLR bounds and intersection bounds sometimes included 0 in covariate groups 3, 4, and 5. After 2009, bounds with or without CLR adjustment do not contain 0 for covariate group 5. These results illustrate that in those periods, women of Type 4 probably had a lower hourly wage than men of the same type. Most of the 95\% confidence intervals include 0 in the later years, while only 95\% CIs from \cite{Imbens2004} do not include 0 after 2010.

Comparing the results for Types 4 and 12, we notice the big difference between the trends of these types. For Type 12, in 1976, the bounds for gender gap included 0 across 4 intervals and 5 covariate groups until 1984 (CLR bounds) or 1982 (intersection bounds). After that, in some years, the lower bounds become positive for covariate group 4. From 2009 onwards, for covariate group 5, the bounds for the ATE and the \cite{Imbens2004} 95\% CIs exclude 0. 

 \section{Conclusion}

This paper addresses a crucial aspect of sample selection within the context of the gender gap problem. In this context, even when treatment assignment is random, the presence of selection bias distorts the treatment's effect on the outcome variable due to unobserved factors. Consequently, bounding the treatment effect without making restrictive distributional or model specification assumptions offers an alternative more robust method. Unlike standard sample selection models, bounding the effect does not necessitate imposing an exclusion restriction. However, employing an exclusion restriction is a common practice in gender gap investigations, with its validity tested in various studies. Utilizing the additional information provided by this exclusion restriction enables the segmentation of the population into distinct subgroups or types, facilitating the derivation of narrower bounds for each type.

Examining the gender gap within the sample selection model and the potential outcome framework allows us to identify the proportions of various types and bound the gender gap for each type accordingly. Notably, we observe distinct trends in the proportion of each type and the corresponding gender gap bounds. For instance, we observe an increasing trend in the proportion of individuals who always work over time, along with varying trends in the bounds for this type. Specifically, we note two key trends. First, the upper and lower bounds of the gender gap tend to shrink over time, albeit at different rates. Prior to the 1990s, the decline was rapid, whereas thereafter, it slowed down. Secondly, we find that the gender gap upper and lower bounds are consistently highest among individuals in the high-potential wage group. With the addition of further assumptions to our basic model, we find that the gender gap bounds for the always-working subgroup show that the gender gap existed in the sample period for this type of individual. However, for other types of individuals, the trend in the gender gap estimates differs or remains ambiguous when examining the 95\% CIs. This underscores the significance of disaggregating individuals by type, as it reveals that the gender gap varies across different potential wage levels and types of worker.

\small
\bibliographystyle{chicago}
\bibliography{references.bib}
\normalsize

\newpage
\appendix

 \section{Tighter Bounds for Types 4 and 12 under Assumption \ref{asu.FOSD.Y0}}

In this section, we provide the tighter bounds for Types 4 and 12 by additionally assuming one dominance assumption. This is done by assuming Assumption \ref{asu.FOSD.Y0} under the basic model. Recall that under the basic model, the bounds of $E(Y^*(0)|\text{Type T})$ with $T \in \{4, 12\}$ are the theoretical bounds. We point identify $E(Y^*(0)|\text{Type 1})$ and $E(Y^*(0)|\text{Type 2})$, so Assumption \ref{asu.FOSD.Y0} is only relevant for Type 4 and 12. Therefore, with Assumption \ref{asu.FOSD.Y0}, the bounds of $E(Y^*(0)|\text{Type T})$ with $T \in \{4, 12\}$ are $[Y^{LB}, \min_{T \in \{1,2\}}  E(Y^*(0)| \text{Type T})]$.

With Assumption \ref{asu.FOSD.Y0}, for Type 4, the bounds on the ATE are as follows: $$UB = \min_{z} E[Y|D = 1, S = 1, Y \geq Y_{1 - q_{T_z}}, Z = z]-Y^{LB}$$
$$LB=  \max_{z} E[Y|D = 1, S = 1, Y \leq Y_{q_{T_z}}, Z = z]- \min_{T \in \{1,2\}}  E(Y^*(0)| \text{Type T})$$

For Type 12, the bounds on the ATE are as follows: $$UB =  E[Y|D = 1, S = 1, Y \geq Y_{1 - q_{T12_0}}, Z = 1]-Y^{LB}$$
$$LB=  E[Y|D = 1, S = 1, Y \leq Y_{q_{T12_0}}, Z = 1]- \min_{T \in \{1,2\}}  E(Y^*(0)| \text{Type T})$$

For the theoretical bounds, we only use the lower bound, that is, $Y^{LB}$. For Type 4, we have a maximum operator and a minimum operator. However, there is a minus sign before the minimum operator, so it is convenient to think that we maximize the negative value of the two terms inside the original minimum operator. That is, we will have two arguments, $Z$ and $T$. We need to calculate the maximum of the term $$E[Y|D = 1, S = 1, Y \leq Y_{q_{T_z}}, Z = z]-E(Y^*(0)| \text{Type T})$$ In this way, the estimation procedure is simplified.
\section{Proofs}
\subsection{Observable Distributions and Unobservable Types}

In this subsection we establish the relationship between observable distributions (or average) of outcome for a cell in Table \ref{TBL:PS.OBSE} and unobservable distributions (or average) for types.
\begin{itemize}
  \item First we show that in the cell $(D = 1, S = 1, Z = 1)$ in Table \ref{TBL:PS.OBSE}, the distribution of outcome variable is a weighted average of distributions of 4 types.
        \begin{align*}
  P_r(Y \leq y &|D = 1, S = 1, Z = 1) * P_r(S = 1|D = 1, Z = 1) \\
   &= P_r(Y^*(1) \leq y|S(1,1) = 1)* P_r(S(1,1) = 1) \\
   &= P_r(Y^*(1) \leq y, S(1,1) = 1)\\
   &= P_r(Y^*(1) \leq y, S(1,1) = 1, S(0,1) = 1) \\
   &+ P_r(Y^*(1) \leq y, S(1,1)  = 1, S(0,1) = 0) \\
   &= P_r(Y^*(1) \leq y|\text{Type 1}) * P_r(\text{Type 1})\\
   &+ \sum_{T \in \{ 2, 4, 12\}} P_r(Y^*(1) \leq y|\text{Type T}) * P_r(\text{Type T})
\end{align*}
    The first equality sign is based on Assumption \ref{asu.indep}, the second is from Bayes' Theorem, the third follows from the additivity of probability for mutually exclusive events, and the last is from Assumptions \ref{asu.indep}-\ref{asu.monoZ}, i.e., the definitions of types. \hfill\qed
  \item Next, we show that observed mean of $Y$ conditional on $(S = 1, D = 0, Z = 0)$ is a weighted average of the mean potential outcome when untreated for Type 1 and Type 2.
        \begin{align*}
 E(Y &|S = 1, D = 0, Z = 0) = E(Y^*(0)|S(0,0) = 1) \\
   &= E(Y^*(0)| S(0,1) = 1, S(0,0) = 1) * \frac{\pi_{T1}}{\pi_{T1}+ \pi_{T2}} \\
   &+ E(Y^*(0)| S(0,1) = 0, S(0,0) = 1)*(1-\frac{\pi_{T1}}{\pi_{T1}+ \pi_{T2}})\\
   &= E(Y^*(0)|\text{Type 1}) * \frac{\pi_{T1}}{\pi_{T1}+ \pi_{T2}} \\
   &+ E(Y^*(0)| \text{Type 2})*\frac{\pi_{T2}}{\pi_{T1}+ \pi_{T2}}
\end{align*}
The first equality sign is from Assumption \ref{asu.indep}, the second is from Bayes' Theorem, Assumptions \ref{asu.indep}-\ref{asu.monoZ}, and the proportions for Type 1 and Type 2, and the last is from Assumptions \ref{asu.indep}-\ref{asu.monoZ}, i.e., the definitions of types.\hfill\qed
\end{itemize}

\subsection{Proof for Asymptotic Properties of the bounds of \texorpdfstring{$E(Y^*(0)|\text{Type T})$}{E(Y(0)|Type T)} for Theorem \ref{theoremCAN}}

In this subsection, we prove the asymptotic distributions for estimators of $E(Y^*(0)|\text{Type T})$. We also provide the formula for $V^{C}$ for Type 1 and Type 2. From now on, we use $V^{C}$ instead of $V^{C}_T$ to simplify the notation.

We apply the GMM method to prove the asymptotic properties of estimators defined in \ref{equ:estT2} and $\widehat{Y_{0,T1}}$. This method is used in deducing the asymptotic normality and estimating the variance of those estimators in \cite{Lee2009} and \cite{ChenXuan2015}. Let $\bar{x}_T$ denote the unknown mean of the untreated for Type T, and $\bar{x}_{12}$ as the unknown representation for $E(Y|S = 1, D = 0, Z = 0)$. The unknown variables are $\theta = (\bar{x}_2,  \bar{x}_1, \bar{x}_{12}, \alpha_1, \alpha_2)$. The corresponding population parameters are $\theta_0 = (E(Y(0)|Type 2), E(Y(0)|Type 1), E(Y|S = 1, D = 0, Z = 0), \pi_{T1}, (\pi_{T1}+\pi_{T2}))$. To simplify the notation, set $\theta_0 = (\mu_{02}, \mu_{01}, \mu_{012}, \pi_{T1}, (\pi_{T1}+\pi_{T2}))$. The moment function is defined as follows:
\begin{equation*}
g_1(.,\theta) = \begin{pmatrix}
   \alpha_2 \bar{x}_{12}  - \alpha_1 \bar{x}_1 - (\alpha_2 - \alpha_1) \bar{x}_2\\
  (Y - \bar{x}_1)S(1-D)Z \\
  (Y - \bar{x}_{12})S(1-D)(1-Z) \\
  (S - \alpha_1)(1- D)Z\\
  (S - \alpha_2)(1- D)(1-Z)\\
\end{pmatrix}
\end{equation*}
 $E(g(.,\theta_0)) = 0$ for all five moments. The first moment is zero by the Law of the Iterated Expectation. The next four moments are zero because of the definition of the population parameters. Since the first equation is always 0, we focus on the following four moments first.

\begin{equation*}
g(.,\theta) = \begin{pmatrix}
  (Y - \bar{x}_1)S(1-D)Z \\
  (Y - \bar{x}_{12})S(1-D)(1-Z) \\
  (S - \alpha_1)(1- D)Z\\
  (S - \alpha_2)(1- D)(1-Z)\\
\end{pmatrix}
\end{equation*}
  
There are four moments and four unknown parameters. For GMM, it is a just-identified case. Following the steps in \cite{Lee2009}, we deduce the asymptotic variance for those two estimators.

The estimators are calculated using the sample version of $E(g(.,\theta_0))$, that is, $$\min_{\theta} (\sum g(.,\theta) )'(\sum g(.,\theta) ).$$ Denote G as the derivative of expected moment function evaluated at the true parameter $\theta_0$ and $\Sigma$ as the asymptotic variance of $\frac{1}{n}\sum g(.,\theta_0)$. The asymptotic variance is $V = G^{-1} \Sigma G'^{-1}$ based on \cite{Newey1994}. G is in the following:
 \begin{equation*}
   \begin{bmatrix}
      -E[S(1-D)Z] & 0 & 0 & 0 \\
      0 & -E[S(1-D)(1-Z)] & 0 & 0 \\
      0 & 0 & -E[(1- D)Z] & 0 \\
      0 & 0 & 0 & -E[(1- D)(1-Z)]
   \end{bmatrix}
 \end{equation*}

After performing the Gauss-Jordan elimination, we obtain the $G^{-1}$ as follows:
 \begin{equation*}
   \begin{bmatrix}
    -\frac{1}{E[S(1-D)Z]} & 0 & 0 & 0 \\
      0 & -\frac{1}{E[S(1-D)(1-Z)]} & 0 & 0 \\
      0 & 0 & -\frac{1}{E[(1- D)Z]} & 0 \\
      0 & 0 & 0 & -\frac{1}{E[(1- D)(1-Z)]}
   \end{bmatrix}
 \end{equation*}

 $\Sigma$ is a $4\times 4$ matrix, i.e.,
 \begin{equation*}
   \begin{bmatrix}
      c_1 & 0 & 0 & 0 \\
      0 & c_2  & 0 & 0 \\
      0 & 0 & \pi_{T1}(1-\pi_{T1})E[(1- D)Z] & 0 \\
      0 & 0 & 0 &  (\pi_{T1}+\pi_{T2})(1-(\pi_{T1}+\pi_{T2})) E[(1- D)(1-Z)]\\
   \end{bmatrix}
 \end{equation*}
 with $c_1 = Var(Y|S = 1, D = 0, Z = 1) E[S(1-D)Z]$ and $c_2 = Var(Y|S = 1, D = 0, Z = 0) E[S(1-D)(1-Z)]$. We obtain the explicit form for the asymptotic variance $G^{-1} \Sigma G'^{-1}$ of the four estimators: $(\hat \mu_{01}, \hat \mu_{012}, \hat \alpha_1, \hat \alpha_2)$. For instance, the asymptotic variance for $\hat \mu_{01}$ is: $$V^C_{T1} =\frac{Var(Y|S = 1, D = 0, Z = 1)}{E[S(1-D)Z]} $$
 
Next, we deduce the asymptotic variance for $\widehat{Y_{0,T2}}$ (or $\hat \mu_{02}$). To simplify the notations, $\alpha_2$ is $(\pi_{T1}+\pi_{T2})$, and $\alpha_1$ is $\pi_{T1}$.
 
 \begin{align*}
   \sqrt{n} (\hat \mu_{02}- \mu_{02}) =& \sqrt{n} \left[ \frac{\hat \alpha_2 \hat \mu_{012} - \hat \alpha_1 \hat \mu_{01}}{\hat \alpha_2 -  \hat \alpha_1}  - \frac{\alpha_2 \mu_{012} - \alpha_1  \mu_{01}}{ \alpha_2 -  \alpha_1} \right]\\
    =& \sqrt{n} \left[ \frac{\hat \alpha_2 \hat \mu_{012} - \hat \alpha_1 \hat \mu_{01}}{\hat \alpha_2 -  \hat \alpha_1} -  \frac{\hat \alpha_2 \mu_{012} - \hat \alpha_1 \mu_{01}}{\hat \alpha_2 -  \hat \alpha_1} + \frac{\hat \alpha_2 \mu_{012} - \hat \alpha_1 \mu_{01}}{\hat \alpha_2 -  \hat \alpha_1}   \right] \\
    &+ \sqrt{n} \left[- \frac{ \alpha_2 \mu_{012} - \alpha_1 \mu_{01}}{\hat \alpha_2 -  \hat \alpha_1} + \frac{ \alpha_2 \mu_{012} - \alpha_1 \mu_{01}}{\hat \alpha_2 -  \hat \alpha_1} - \frac{\alpha_2 \mu_{012} - \alpha_1  \mu_{01}}{ \alpha_2 -  \alpha_1} \right]\\
    =& \frac{1}{\hat \alpha_2 -  \hat \alpha_1} \left[ \hat \alpha_2 \sqrt{n}(\hat \mu_{012} - \mu_{012}) - \hat \alpha_1 \sqrt{n}(\hat \mu_{01} - \mu_{01})  \right]\\
    &+ \frac{1}{\hat \alpha_2 -  \hat \alpha_1} \left[ \mu_{012} \sqrt{n}(\hat \alpha_2 - \alpha_2) - \mu_{01} \sqrt{n}(\hat \alpha_1 - \alpha_1)  \right]\\
    &+ \frac{\alpha_2 \mu_{012} - \alpha_1  \mu_{01}}{ \alpha_2 -  \alpha_1} \left[ \sqrt{n} (\frac{\alpha_2 -  \alpha_1}{\hat \alpha_2 -  \hat \alpha_1} -1)\right]\\
    =& \frac{1}{\hat \alpha_2 -  \hat \alpha_1} \left[ \hat \alpha_2 \sqrt{n}(\hat \mu_{012} - \mu_{012}) - \hat \alpha_1 \sqrt{n}(\hat \mu_{01} - \mu_{01})  \right]\\
    &+ \frac{1}{\hat \alpha_2 -  \hat \alpha_1} \left[(\mu_{012} - \mu_{02}) \sqrt{n}(\hat \alpha_2 - \alpha_2) - (\mu_{01} -  \mu_{02})\sqrt{n}(\hat \alpha_1 - \alpha_1)  \right]\\
 \end{align*}
 
According to the Delta Method, the asymptotic variance of $\hat \mu_{02}$ is a formula with four components, as shown in the following:
\begin{align*}
 V^C_{T2} &= \frac{\pi_{T1}^2 Var(Y|S = 1, D = 0, Z = 1)}{\pi_{T2}^2 E[S(1-D)Z]} + \frac{(\pi_{T1}+ \pi_{T2})^2 Var(Y|S = 1, D = 0, Z = 0)}{\pi_{T2}^2 E[S(1-D)(1-Z)]} \\
   &+ \frac{\pi_{T1}(1-\pi_{T1})(E(Y(0)|Type 2) - E(Y(0)|Type 1))^2}{\pi_{T2}^2 E[(1-D)Z]} \\
   &+ \frac{(\pi_{T1}+ \pi_{T2})(1-(\pi_{T1}+ \pi_{T2}))(E(Y(0)|Type 2) - E(Y|S = 1, D = 0, Z = 0))^2}{\pi_{T2}^2 E[(1-D)(1-Z)]} \\
 Cov(\hat \mu_{02}, \hat \mu_{01})  &=  \frac{\pi_{T1} Var(Y|S = 1, D = 0, Z = 1)}{\pi_{T2} E[S(1-D)Z]}
\end{align*}

The first two terms of $V^C_{T2}$ incorporate the variance from estimating $E(Y(0)|Type 1)$ and $E(Y|S = 1, D = 0, Z = 0)$ and the last two contain the variance of estimating $\pi_{T1}$ and $(\pi_{T1}+\pi_{T2})$.

 For Type 4 and Type 12, because we are using theoretical bounds and the minimum or maximum of the outcome variable as the estimator, we use the bootstrap method to find the corresponding standard error for the estimator. \footnote{We thank Xuan Chen for her insights on finding the estimator and the standard error.}
\subsection{Proof for Asymptotic Properties of the bounds of \texorpdfstring{$E(Y^*(1)|\text{Type T})$}{E(Y(1)|Type T)} for Theorem \ref{theoremCAN}}

In this subsection, we provide the asymptotic distribution for estimators of upper and lower bounds of $E(Y^*(1)|\text{Type T})$. That is, we provide the formula for $V^{LB}_T$ and $V^{UB}_T$ for each type and each $Z$. From now on, we use $V^{LB}$ and $V^{UB}$ to simplify the notation.

For each type, the formulas for estimators and their asymptotic distributions are different. Here are proofs for each type. When we consider the cases $Z = 1$ and $Z = 0$ separately, there are four moments for Type 1 and five moments for other types. The first moment contains the key parameter for the bounds. The last moment (for Type 1) or the last two moments (for the other types) give us (part of) the proportion of each type in the population. The third one delivers the proportion of this type in the case ($D = 1, Z  = z$). The second one produces the quantile $y_p$. The first one generates $\mu^{LB}$. Those are the key parameters we want to estimate. The proof in this subsection is similar to the one in the previous section. We construct the moments and use the GMM method as in \cite{Lee2009}. We only provide the explicit expressions for $G$, $G^{-1}$, and $\Sigma$ for $\widehat{LB}$ for Type 1 when $Z = 1$, because the methodology of calculating $G$, $G^{-1}$, and $\Sigma$ for other types and for $\widehat{UB}$ is similar.

\begin{itemize}
  \item Type 1
\end{itemize}
For Type 1, when $Z = 1$, there are 4 moments with 4 parameters to estimate. The key parameter is $LB$ with the notation $\mu^{LB}$. The other three parameters are: $y_{p_0}$ (the quantile), $p_0$ (the proportion of type 1 when $D = 1$ and $Z  = 1$), and $\alpha_0$ (the proportion of Type 1). To simplify the notations, we use $(\mu_0, y_{p_0}, p_0, \alpha_0)$ correspondingly.
$$\mu^{LB} = E[Y|D = 1, S = 1, Y \leq y_{p_0}, Z = 1]$$
\begin{equation*}
E\begin{pmatrix}
  (Y - \mu^{LB} ) 1[Y < y_{p_0}]SDZ \\
  (1[Y > y_{p_0}] - 1 + p_0) SDZ \\
  (S - \alpha_0 \frac{1}{p_0})DZ \\
  (S - \alpha_0)(1- D)Z\\
\end{pmatrix} = 0
\end{equation*}
Based on the 4 moments, we define $g(., \theta)$ in the following:
\begin{equation*}
g(., \theta) = \begin{pmatrix}
  (Y - \mu) 1[Y < y_p]SDZ \\
  (1[Y > y_p] - 1 + p) SDZ \\
  (S - \alpha \frac{1}{p})DZ \\
  (S - \alpha)(1- D)Z\\
\end{pmatrix}
\end{equation*}
The 4 estimators proposed in \ref{sec:estimator} are the sample analogues of the parameters that satisfy those moments. Define G as the derivative of $E[g(., \theta)]$ when $\theta = \theta_0$ and $\Sigma$ as the asymptotic variance of $n^{-1} \Sigma g(.,\theta_0)$. Then the asymptotic variance for all of the estimators is $V = G^{-1} \Sigma G'^{-1}$ based on \cite{Newey1994}. Here, we adopt similar notations and procedures as in \cite{Lee2009}. Check the details in the proof of Proposition 3 in \cite{Lee2009}.

G is in the following:
\begin{equation*}
  \begin{bmatrix}
    -E[SDZ]p_0 & f(y_{p_0})E[SDZ](y_{p_0} - \mu_0) & 0 & 0 \\
    0 & -f(y_{p_0})E[SDZ] & E[SDZ] & 0 \\
    0 & 0 & -\frac{\alpha_0}{p^2_0}E[DZ] & -\frac{1}{p_0} E[DZ] \\
    0 & 0 & 0 & -E[(1- D)Z]
  \end{bmatrix}
\end{equation*}
with $f(.)$ the PDF of Y conditional on $(S =1, D= 1, Z = 1)$.
After performing the Gauss-Jordan elimination, we obtain the $G^{-1}$ as follows:
\begin{equation*}
  \begin{bmatrix}
     \frac{1}{-E[SDZ]P_0} &  -\frac{(y_{p_0} - \mu_0)}{E[SDZ]p_0} &-\frac{p_0(y_{p_0} - \mu_0)}{\alpha_0E[DZ]} & \frac{1}{E[(1- D)Z]}  \frac{(y_{p_0} - \mu_0)}{\alpha_0}   \\
    0 & -\frac{1}{f(y_{p_0})E[SDZ]} & -\frac{p^2_0}{\alpha_0E[DZ]f(y_{p_0})} & \frac{1}{E[(1- D)Z]} \frac{1}{f(y_{p_0})} \frac{p_0}{\alpha_0}\\
    0 & 0 & -\frac{p^2_0}{\alpha_0E[DZ]} & \frac{1}{E[(1- D)Z]} \frac{p_0}{\alpha_0}\\
    0 & 0 & 0 & -\frac{1}{E[(1- D)Z]}
  \end{bmatrix}
\end{equation*}
$\Sigma$ is a $4\times 4$ matrix, i.e.,
\begin{equation*}
  \begin{bmatrix}
      \int_{-\infty}^{y_{p_0}} (y - \mu_0)^2 f(y) dyE[SDZ] & 0 & 0 & 0 \\
    0 &  p_0(1-p_0)E[SDZ]   & 0 & 0 \\
    0 & 0  &  \frac{\alpha_0}{p_0}(1-\frac{\alpha_0}{p_0}) E[DZ] & 0 \\
    0 & 0 & 0 &  \alpha_0 (1-\alpha_0) E[(1-D)Z]
  \end{bmatrix}
\end{equation*}

The asymptotic variance for the first parameter $\mu^{LB}$ is the element in the first row and first column of $G^{-1} \Sigma G'^{-1}$:
\begin{align*}
V^{LB}&= \frac{Var(Y|S = 1, D = 1, Z = 1,Y \leq y_{p_0} )}{E(SDZ) p_0} + \frac{(y_{p_0} - \mu^{LB})^2(1-p_0)}{E(SDZ) p_0} \\
   &+ \frac{(y_{p_0} - \mu^{LB})^2}{(p_0)^2} V_p \\
 V_p &= (p_0)^2 [\frac{1-\frac{\alpha_0}{p_0}}{\frac{\alpha_0}{p_0} E(DZ)} + \frac{(1-\alpha_0)}{\alpha_0 E(Z[1-D])} ]
\end{align*}

Next, we find the asymptotic distribution for the estimator of $\mu^{LB}  = E[Y|D = 1, S = 1, Y \leq y_{p}, Z = 0]$ when $Z = 0$ using the moment function $g(., \theta)$ in the following:
\begin{equation*}
g(., \theta) = \begin{pmatrix}
  (Y - \mu) 1[Y < y_p]SD(1-Z) \\
  (1[Y > y_p] - 1 + p) SD(1-Z) \\
  (S - \alpha \frac{1}{p})D(1-Z) \\
  (S - \alpha)(1- D)Z\\
\end{pmatrix}
\end{equation*}
 The asymptotic variance for the estimator of $\mu^{LB}$ is:
\begin{align*}
V^{LB}&= \frac{Var(Y|S = 1, D = 1, Z = 0,Y \leq y_{p_0})}{E(SD(1-Z)) p_0} + \frac{(y_{p_0} - \mu^{LB})^2(1-p_0)}{E(SD(1-Z)) p_0} \\
   &+ \frac{(y_{p_0} - \mu^{LB})^2}{(p_0)^2} V_p \\
 V_p &= (p_0)^2 [\frac{1-\frac{\alpha_0}{p_0}}{\frac{\alpha_0}{p_0} E(D(1-Z))} + \frac{(1-\alpha_0)}{\alpha_0 E(Z[1-D])} ]
\end{align*}
with $p_0$ the proportion of type 1 in the case ($D = 1, Z  = 0$).

Alternatively, to estimate the true parameters for $Z = 0$ and $Z = 1$ all together, we construct the moment function $g(., \theta)$ as:
\begin{equation*}
g(., \theta) = \begin{pmatrix}
  (Y - \mu) 1[Y < y_p]SDZ \\
  (Y - \mu') 1[Y < y_p']SD(1-Z) \\
  (1[Y > y_p] - 1 + p) SDZ \\
  (1[Y > y_p'] - 1 + p') SD(1-Z) \\
  (S - \alpha \frac{1}{p})DZ \\
  (S - \alpha \frac{1}{p'})D(1-Z) \\
  (S - \alpha)(1- D)Z\\
\end{pmatrix}
\end{equation*}
Using the moments to construct the asymptotic variance matrix, we obtain the asymptotic variances for $\widehat{LB}$ when $Z =1$ and $Z = 0$ at the same time. The covariance between those two $\widehat{LB}$s is: $$\frac{(1-\alpha_0)[y_{p_{0,Z=1}} - \mu^{LB}_{Z=1}][y_{p_0, Z=0} - \mu^{LB}_{Z=0}]}{\alpha_0 E(Z[1-D])}.$$

\begin{itemize}
  \item Type 2
\end{itemize}
For Type 2, we follow the same steps as before, except that there are 5 parameters to estimate.

When $Z = 1$ with $\mu^{LB}= E[Y|D = 1, S = 1, Y \leq y_{p}, Z = 1]$, we construct the moment function $g(., \theta)$ as:
\begin{equation*}
g(., \theta) = \begin{pmatrix}
  (Y - \mu) 1[Y < y_p]SDZ \\
  (1[Y > y_p] - 1 + p) SDZ \\
  (S - (\alpha_1 - \alpha_2) \frac{1}{p})DZ \\
  (S - \alpha_1) (1-D)(1-Z)\\
  (S - \alpha_2) (1-D)Z
\end{pmatrix}
\end{equation*}
 The asymptotic variance for the estimator of $\mu^{LB}$ is
\begin{align*}
V^{LB} &= \frac{Var(Y|S = 1, D = 1, Z = 1,Y \leq y_{p} )}{E(SDZ) p_0} + \frac{(y_{p_0} - \mu^{LB})^2(1-p_0)}{E(SDZ) p_0} \\
   &+ \frac{(y_{p_0} - \mu^{LB})^2}{(p_0)^2} V_p \\
 V_p &= (p_0)^2 [\frac{1-\frac{\alpha_1 - \alpha_2}{p_0}}{\frac{\alpha_1 - \alpha_2}{p_0} E(DZ)} + \frac{\alpha_1(1-\alpha_1)}{(\alpha_1 - \alpha_2)^2 E([1-D](1-Z))} + \frac{\alpha_2(1-\alpha_2)}{(\alpha_1- \alpha_2)^2 E([1-D]Z)}]
\end{align*}
with every parameters under condition $Z = 1$.

When $Z = 0$ with $\mu^{LB} = E[Y|D = 1, S = 1, Y \leq y_{p}, Z = 0]$, the moment function is:
\begin{equation*}
g(., \theta) = \begin{pmatrix}
  (Y - \mu) 1[Y < y_p]SD(1-Z) \\
  (1[Y > y_p] - 1 + p) SD(1-Z) \\
  (S - (\alpha_1 - \alpha_2) \frac{1}{p})D(1-Z) \\
  (S - \alpha_1) (1-D)(1-Z)\\
  (S - \alpha_2) (1-D)Z
\end{pmatrix}
\end{equation*}
 The asymptotic variance for the estimator of $\mu^{LB}$ when $Z  = 0$ is
\begin{align*}
V^{LB} &= \frac{Var(Y|S = 1, D = 1, Z = 0,Y \leq y_{p} )}{E(SD(1-Z)) p_0} + \frac{(y_{p_0} - \mu^{LB})^2(1-p_0)}{E(SD(1-Z)) p_0} \\
   &+ \frac{(y_{p_0} - \mu^{LB})^2}{(p_0)^2} V_p \\
 V_p &= (p_0)^2 [\frac{1-\frac{\alpha_1 - \alpha_2}{p_0}}{\frac{\alpha_1 - \alpha_2}{p_0} E(D(1-Z))} + \frac{\alpha_1(1-\alpha_1)}{(\alpha_1 - \alpha_2)^2 E([1-D](1-Z))} + \frac{\alpha_2(1-\alpha_2)}{(\alpha_1- \alpha_2)^2 E([1-D]Z)}]
\end{align*}

If we want to estimate the parameters for $Z=0$ and $Z = 1$ all together, the moment function becomes:

\begin{equation*}
g(., \theta) = \begin{pmatrix}
  (Y - \mu) 1[Y < y_p]SDZ \\
   (Y - \mu') 1[Y < y_p']SD(1-Z) \\
  (1[Y > y_p] - 1 + p) SDZ \\
  (1[Y > y_p'] - 1 + p') SD(1-Z) \\
  (S - (\alpha_1 - \alpha_2) \frac{1}{p})DZ \\
  (S - (\alpha_1 - \alpha_2) \frac{1}{p'})D(1-Z) \\
  (S - \alpha_1) (1-D)(1-Z)\\
  (S - \alpha_2) (1-D)Z
\end{pmatrix}
\end{equation*}

Using the moments to construct the asymptotic variance matrix, we obtain the asymptotic variances for $\widehat{LB}$ when $Z =1$ and $Z = 0$ at the same time. The covariance between those two $\widehat{LB}$s is: $$\frac{\alpha_1(1-\alpha_1)(y_{p_0, Z =1} - \mu^{LB}_{Z = 1})(y_{p_0, Z= 0} - \mu^{LB}_{Z=0})}{(\alpha_1 - \alpha_2)^2 E([1-D](1-Z))} + \frac{\alpha_2(1-\alpha_2)(y_{p_0, Z =1} - \mu^{LB}_{Z = 1})(y_{p_0, Z= 0} - \mu^{LB}_{Z=0})}{(\alpha_1- \alpha_2)^2 E([1-D]Z)}$$

\begin{itemize}
  \item Type 4
\end{itemize}

For Type 4, there are 5 unknown parameters to estimate. When $Z = 1$ with $\mu^{LB} = E[Y|D = 1, S = 1, Y \leq y_{p}, Z = 1]$,  the moment function is:
\begin{equation*}
g(., \theta) = \begin{pmatrix}
  (Y - \mu) 1[Y < y_p]SDZ \\
  (1[Y > y_p] - 1 + p) SDZ \\
  (S - (\alpha_1 - \alpha_2) \frac{1}{p})DZ \\
  (S - \alpha_1) D(1-Z)\\
  (S - \alpha_2) (1-D)(1-Z)
\end{pmatrix}
\end{equation*}
 The asymptotic variance for the estimator of $\mu^{LB}$ when $Z  = 1$ is:
\begin{align*}
V^{LB} &= \frac{Var(Y|S = 1, D = 1, Z = 1,Y \leq y_{p} )}{E(SDZ) p_0} + \frac{(y_{p_0} - \mu^{LB})^2(1-p_0)}{E(SDZ) p_0} \\
   &+ \frac{(y_{p_0} - \mu^{LB})^2}{(p_0)^2} V_p \\
 V_p &= (p_0)^2 [\frac{1-\frac{\alpha_1 - \alpha_2}{p_0}}{\frac{\alpha_1 - \alpha_2}{p_0} E(DZ)} + \frac{\alpha_1(1-\alpha_1)}{(\alpha_1 - \alpha_2)^2 E(D(1-Z))} + \frac{\alpha_2(1-\alpha_2)}{(\alpha_1- \alpha_2)^2 E([1-D](1-Z))}]
\end{align*}

Next, when $Z = 0$, $\mu^{LB} = E[Y|D = 1, S = 1, Y \leq y_{p}, Z = 0]$.
\begin{equation*}
g(., \theta) = \begin{pmatrix}
  (Y - \mu) 1[Y < y_p]SD(1-Z) \\
  (1[Y > y_p] - 1 + p) SD(1-Z) \\
  (S - (\alpha_1 - \alpha_2) \frac{1}{p})D(1-Z) \\
  (S - \alpha_1) D(1-Z)\\
  (S - \alpha_2) (1-D)(1-Z)
\end{pmatrix}
\end{equation*}
 The asymptotic variance for the estimator of $\mu^{LB}$ when $Z  = 0$ is:
\begin{align*}
V^{LB} &= \frac{Var(Y|S = 1, D = 1, Z = 0,Y \leq y_{p} )}{E(SD(1-Z)) p_0} + \frac{(y_{p_0} - \mu^{LB})^2(1-p_0)}{E(SD(1-Z)) p_0} \\
   &+ \frac{(y_{p_0} - \mu^{LB})^2}{(p_0)^2} V_p \\
 V_p &= (p_0)^2 [\frac{1-\frac{\alpha_1 - \alpha_2}{p_0}}{\frac{\alpha_1 - \alpha_2}{p_0} E(D(1-Z))} + \frac{\alpha_1(1-\alpha_1)}{(\alpha_1 - \alpha_2)^2 E(D(1-Z))} + \frac{\alpha_2(1-\alpha_2)}{(\alpha_1- \alpha_2)^2 E([1-D](1-Z))}]
\end{align*}
If we want to estimate the parameters for $Z=0$ and $Z = 1$ all together, the moment function becomes:
\begin{equation*}
g(., \theta) = \begin{pmatrix}
  (Y - \mu) 1[Y < y_p]SDZ \\
  (Y - \mu') 1[Y < y_p']SD(1-Z) \\
  (1[Y > y_p] - 1 + p) SDZ \\
   (1[Y > y_p'] - 1 + p') SD(1-Z) \\
  (S - (\alpha_1 - \alpha_2) \frac{1}{p})DZ \\
 (S - (\alpha_1 - \alpha_2) \frac{1}{p'})D(1-Z) \\
  (S - \alpha_1) D(1-Z)\\
  (S - \alpha_2) (1-D)(1-Z)
\end{pmatrix}
\end{equation*}
Using the moments to construct the asymptotic variance matrix, we obtain the asymptotic variances for $\widehat{LB}$ when $Z =1$ and $Z = 0$ at the same time. The covariance between those two $\widehat{LB}$s is:
$$\frac{\alpha_1(1-\alpha_1)(y_{p_0, Z =1} - \mu^{LB}_{Z = 1})(y_{p_0, Z= 0} - \mu^{LB}_{Z=0})}{(\alpha_1 - \alpha_2)^2 E(D(1-Z))} + \frac{\alpha_2(1-\alpha_2)(y_{p_0, Z =1} - \mu^{LB}_{Z = 1})(y_{p_0, Z= 0} - \mu^{LB}_{Z=0})}{(\alpha_1- \alpha_2)^2 E([1-D](1-Z))}$$

\begin{itemize}
  \item Type 12
\end{itemize}
For Type 12, as before, we focus on the estimator of $\mu^{LB} = E[Y|D = 1, S = 1, Y \leq y_{p}, Z = 1]$. The moment function becomes:
\begin{equation*}
g(., \theta) = \begin{pmatrix}
  (Y - \mu) 1[Y < y_p]SDZ \\
  (1[Y > y_p] - 1 + p) SDZ \\
  (S - (\alpha_1 - \alpha_2) \frac{1}{p})DZ \\
  (S - \alpha_1) DZ\\
  (S - \alpha_2) D(1-Z)
\end{pmatrix}
\end{equation*}
 The asymptotic variance for the estimator of $\mu^{LB}$ when $Z  = 1$ is:
\begin{align*}
V^{LB} &= \frac{Var(Y|S = 1, D = 1, Z = 1,Y \leq y_{p} )}{E(SDZ) p_0} + \frac{(y_{p_0} - \mu^{LB})^2(1-p_0)}{E(SDZ) p_0} \\
   &+ \frac{(y_{p_0} - \mu^{LB})^2}{(p_0)^2} V_p \\
 V_p &= (p_0)^2 [\frac{1-\frac{\alpha_1 - \alpha_2}{p_0}}{\frac{\alpha_1 - \alpha_2}{p_0} E(DZ)} + \frac{\alpha_1(1-\alpha_1)}{(\alpha_1 - \alpha_2)^2 E(ZD)} + \frac{\alpha_2(1-\alpha_2)}{(\alpha_1- \alpha_2)^2 E(D(1-Z))}]
\end{align*}

\section{Empirical Results}

\begin{figure}[!htbp]
\caption{Upper and lower CLR bounds of the Gender Gap for Type 1 and Type 2 across years and 5 covariate groups under basic model} \label{fig:type1and2}
  \centering
\includegraphics[width=0.9\linewidth]{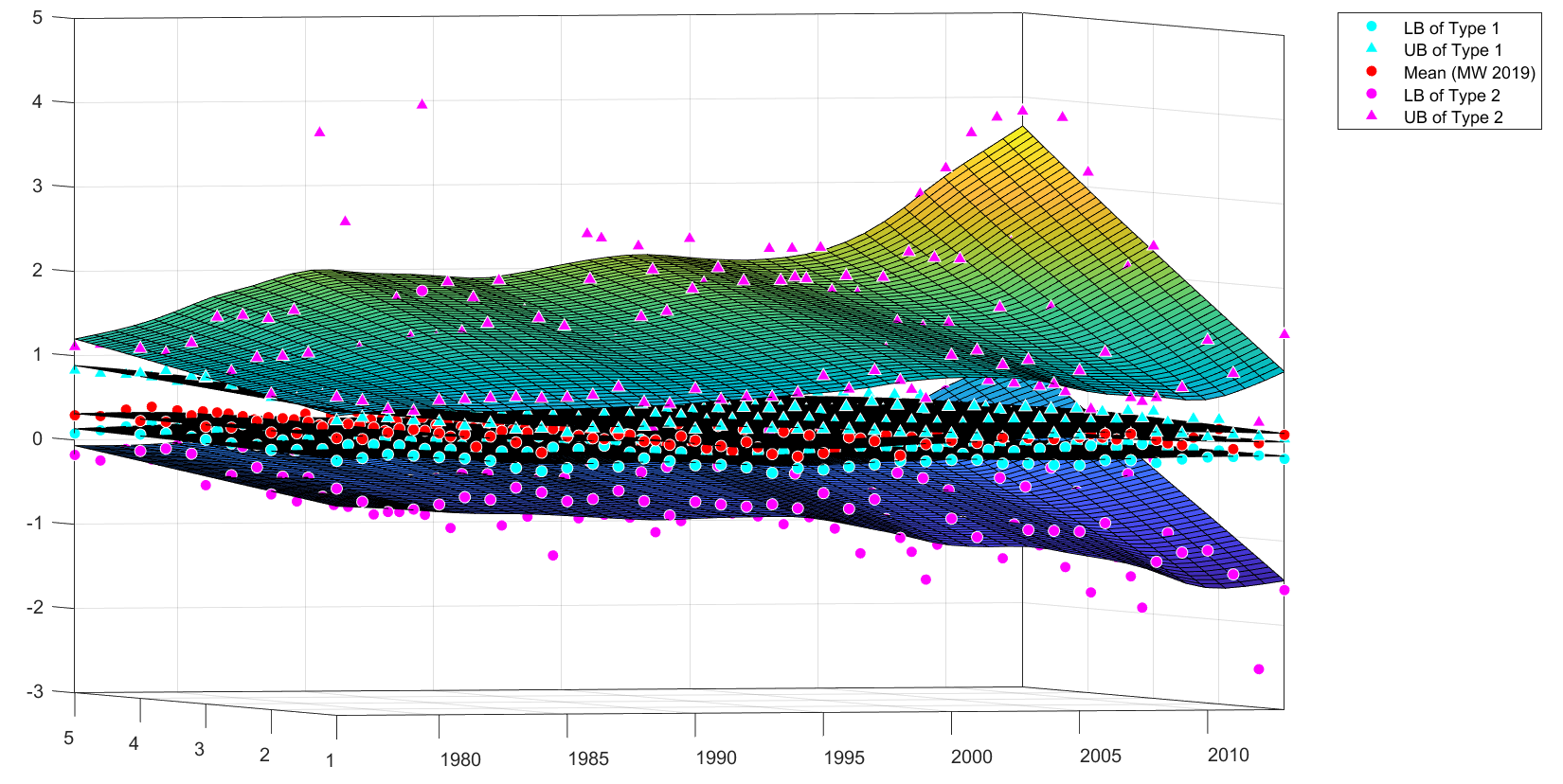}
\end{figure}

In this section, we present additional results for Section \ref{sec:emp}. Firstly, we display the CLR bounds for Type 1 and Type 2 across years and five covariate groups in Figure \ref{fig:type1and2}. The label "Mean (MW 2019)" refers to the gender gap results from \cite{Maasoumi2019} (Table 10, first column: Mean). Notably, Figure \ref{fig:type1and2} illustrates that the bounds of ATE for Type 2 are considerably wide. Specifically, for covariate group 5, the lower bound even surpasses the mean results from \cite{Maasoumi2019} at the end of the sample period.

\begin{figure}[!htbp]
\caption{Upper and lower bounds of the Gender Gap for Type 1 and Type 2 average across covariate groups under basic model}\label{fig:type1and2avg}
  \centering
\includegraphics[width=0.9\linewidth]{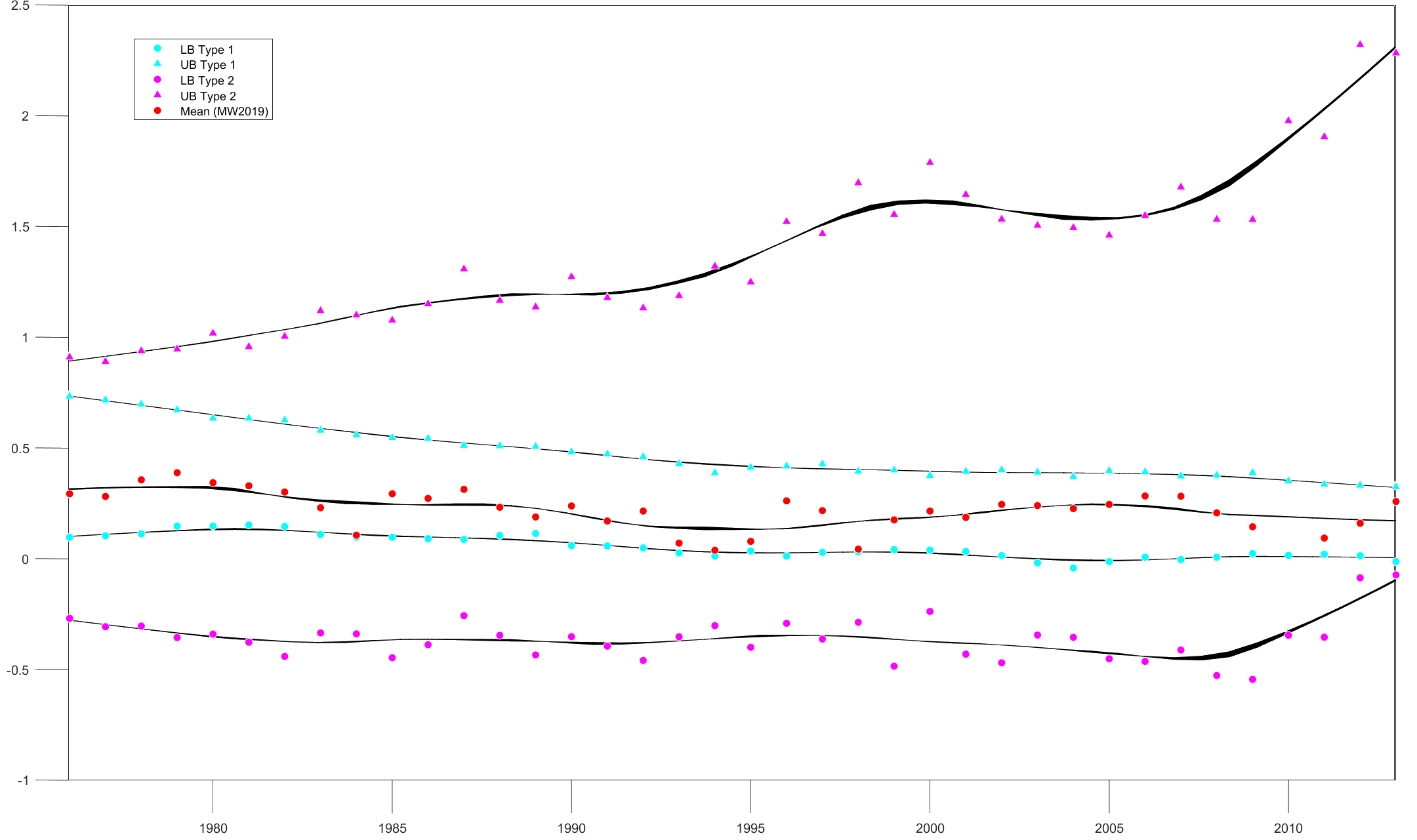}
\end{figure}

If we calculate the average bounds across the five covariate groups, we observe that the bounds for $E(Y^*(0)|\text{Type 2})$ still encompass the mean results from \cite{Maasoumi2019}. This is depicted in Figure \ref{fig:type1and2avg}. The aggregate bounds are computed using the intersection bounds instead of the CLR bounds, as there is no algorithm to calculate the average bounds over all covariate groups with CLR adjustments. While we cannot derive the aggregate bounds using CLR bounds, we anticipate that they would be wider because the CLR bounds are wider than the intersection bounds. Consequently, they would also encompass the mean results from \cite{Maasoumi2019}.

Table \ref{table:T1T2y0} presents the 95\% CIs of the average $E(Y^*(0)|\text{Type T})$ with $T \in {1, 2}$ across all covariate groups over the years. We adopt the method proposed by \cite{Lee2009} to compute the average over covariate groups, which involves applying the iterated expectation theorem.
 \begin{table}[!htbp]
  \centering
   \scalebox{.75}{
\begin{tabular}{ccc|ccc}
\hline
\hline
year & Type 1 & Type 2 & year & Type 1 & Type 2\\ \hline
1976 &[ 2.238, 2.276]&[ 2.287, 2.393]&	1995 &[ 2.382, 2.415]&[ 1.908, 2.202]\\
1977 &[ 2.255, 2.289]&[ 2.318, 2.416]&	1996 &[ 2.389, 2.424]&[ 1.834, 2.175]\\
1978 &[ 2.272, 2.306]&[ 2.280, 2.396]&	1997 &[ 2.387, 2.423]&[ 1.956, 2.275]\\
1979 &[ 2.283, 2.318]&[ 2.263, 2.412]&	1998 &[ 2.412, 2.447]&[ 1.849, 2.280]\\
1980 &[ 2.294, 2.325]&[ 2.217, 2.368]&	1999 &[ 2.422, 2.459]&[ 2.058, 2.504]\\
1981 &[ 2.251, 2.280]&[ 2.227, 2.370]&	2000 &[ 2.457, 2.493]&[ 1.788, 2.267]\\
1982 &[ 2.248, 2.280]&[ 2.113, 2.307]&	2001 &[ 2.464, 2.491]&[ 2.128, 2.436]\\
1983 &[ 2.285, 2.318]&[ 1.813, 2.274]&	2002 &[ 2.477, 2.505]&[ 2.253, 2.524]\\
1984 &[ 2.306, 2.338]&[ 1.895, 2.185]&	2003 &[ 2.505, 2.533]&[ 2.176, 2.402]\\
1985 &[ 2.298, 2.330]&[ 2.078, 2.279]&	2004 &[ 2.524, 2.553]&[ 2.104, 2.341]\\
1986 &[ 2.320, 2.352]&[ 1.993, 2.211]&	2005 &[ 2.495, 2.524]&[ 2.161, 2.407]\\
1987 &[ 2.354, 2.387]&[ 1.733, 2.295]&	2006 &[ 2.482, 2.509]&[ 2.120, 2.415]\\
1988 &[ 2.359, 2.390]&[ 2.013, 2.237]&	2007 &[ 2.493, 2.521]&[ 2.078, 2.436]\\
1989 &[ 2.359, 2.390]&[ 2.064, 2.307]&	2008 &[ 2.497, 2.525]&[ 2.144, 2.500]\\
1990 &[ 2.379, 2.409]&[ 1.962, 2.207]&	2009 &[ 2.477, 2.506]&[ 2.089, 2.430]\\
1991 &[ 2.361, 2.390]&[ 2.017, 2.257]&	2010 &[ 2.512, 2.540]&[ 1.789, 2.320]\\
1992 &[ 2.361, 2.391]&[ 2.053, 2.264]&	2011 &[ 2.506, 2.534]&[ 1.641, 2.283]\\
1993 &[ 2.385, 2.416]&[ 1.950, 2.174]&	2012 &[ 2.500, 2.530]&[ 1.144, 2.278]\\
1994 &[ 2.400, 2.432]&[ 1.814, 2.130]&	2013 &[ 2.505, 2.536]&[ 1.228, 2.108]\\
 \hline
\hline
\end{tabular}
}
\caption{95\% CIs of average $E(Y^*(0)|\text{Type T})$ over all covariate groups across the years}\label{table:T1T2y0}
\end{table}

\begin{table}
  \centering
   \scalebox{.75}{
\begin{tabular}{cccccc}
\hline
\hline
Covariate Group & 1 &2 &3 &4 &5 \\ \hline
Year &[ LB, UB]&[ LB, UB]&[ LB, UB]&[ LB, UB]&[ LB, UB]\\
\hline
1976 &[ 0.025, 0.499]&[ 0.087, 0.706]&[ 0.140, 0.880]&[ 0.133, 0.863]&[ 0.077, 0.826]\\
1977 &[ 0.057, 0.536]&[ 0.085, 0.684]&[ 0.094, 0.782]&[ 0.146, 0.888]&[ 0.113, 0.788]\\
1978 &[ 0.096, 0.530]&[ 0.093, 0.688]&[ 0.076, 0.753]&[ 0.102, 0.822]&[ 0.157, 0.772]\\
1979 &[ 0.081, 0.497]&[ 0.140, 0.641]&[ 0.127, 0.752]&[ 0.168, 0.799]&[ 0.189, 0.740]\\
1980 &[ 0.061, 0.479]&[ 0.141, 0.600]&[ 0.134, 0.694]&[ 0.248, 0.763]&[ 0.132, 0.685]\\
1981 &[ 0.042, 0.429]&[ 0.130, 0.609]&[ 0.196, 0.731]&[ 0.194, 0.725]&[ 0.170, 0.707]\\
1982 &[ 0.014, 0.455]&[ 0.086, 0.605]&[ 0.144, 0.706]&[ 0.219, 0.721]&[ 0.223, 0.667]\\
1983 &[ -0.072, 0.386]&[ 0.077, 0.567]&[ 0.138, 0.668]&[ 0.165, 0.630]&[ 0.179, 0.652]\\
1984 &[ -0.114, 0.393]&[ 0.033, 0.557]&[ 0.113, 0.621]&[ 0.151, 0.630]&[ 0.222, 0.604]\\
1985 &[ -0.079, 0.342]&[ 0.051, 0.499]&[ 0.111, 0.611]&[ 0.173, 0.684]&[ 0.181, 0.616]\\
1986 &[ -0.031, 0.392]&[ 0.040, 0.497]&[ 0.056, 0.565]&[ 0.123, 0.639]&[ 0.205, 0.630]\\
1987 &[ -0.064, 0.380]&[ 0.093, 0.477]&[ 0.107, 0.574]&[ 0.111, 0.528]&[ 0.139, 0.618]\\
1988 &[ 0.030, 0.409]&[ 0.072, 0.438]&[ 0.116, 0.586]&[ 0.123, 0.578]&[ 0.143, 0.557]\\
1989 &[ -0.005, 0.443]&[ 0.111, 0.486]&[ 0.150, 0.571]&[ 0.155, 0.535]&[ 0.112, 0.524]\\
1990 &[ -0.056, 0.370]&[ 0.029, 0.464]&[ 0.107, 0.590]&[ 0.073, 0.500]&[ 0.101, 0.515]\\
1991 &[ -0.052, 0.404]&[ 0.048, 0.476]&[ 0.120, 0.529]&[ 0.059, 0.481]&[ 0.072, 0.496]\\
1992 &[ -0.085, 0.363]&[ 0.027, 0.458]&[ 0.095, 0.546]&[ 0.048, 0.491]&[ 0.107, 0.465]\\
1993 &[ -0.152, 0.372]&[ -0.008, 0.397]&[ 0.108, 0.525]&[ 0.045, 0.461]&[ 0.074, 0.415]\\
1994 &[ -0.097, 0.360]&[ 0.002, 0.413]&[ 0.030, 0.453]&[ -0.012, 0.349]&[ 0.081, 0.399]\\
1995 &[ -0.118, 0.362]&[ 0.011, 0.407]&[ 0.069, 0.442]&[ 0.098, 0.438]&[ 0.051, 0.437]\\
1996 &[ -0.084, 0.353]&[ 0.018, 0.415]&[ 0.054, 0.473]&[ 0.025, 0.412]&[ 0.000, 0.467]\\
1997 &[ -0.061, 0.340]&[ 0.019, 0.412]&[ 0.054, 0.478]&[ 0.071, 0.474]&[ 0.017, 0.460]\\
1998 &[ -0.019, 0.317]&[ 0.038, 0.374]&[ 0.075, 0.469]&[ -0.009, 0.392]&[ 0.037, 0.455]\\
1999 &[ -0.028, 0.311]&[ 0.076, 0.387]&[ 0.107, 0.483]&[ 0.017, 0.386]&[ 0.000, 0.473]\\
2000 &[ -0.012, 0.291]&[ 0.070, 0.369]&[ 0.068, 0.428]&[ 0.090, 0.423]&[ -0.057, 0.396]\\
2001 &[ -0.015, 0.263]&[ 0.063, 0.375]&[ 0.079, 0.480]&[ 0.037, 0.403]&[ -0.026, 0.475]\\
2002 &[ -0.059, 0.244]&[ 0.060, 0.354]&[ 0.067, 0.540]&[ 0.004, 0.369]&[ -0.026, 0.529]\\
2003 &[ -0.063, 0.277]&[ -0.014, 0.342]&[ 0.030, 0.491]&[ -0.034, 0.361]&[ -0.037, 0.511]\\
2004 &[ -0.082, 0.288]&[ 0.031, 0.380]&[ -0.015, 0.445]&[ -0.071, 0.347]&[ -0.097, 0.420]\\
2005 &[ -0.086, 0.267]&[ 0.014, 0.388]&[ 0.039, 0.473]&[ -0.055, 0.362]&[ -0.004, 0.510]\\
2006 &[ -0.030, 0.274]&[ 0.040, 0.400]&[ 0.028, 0.478]&[ -0.011, 0.326]&[ -0.017, 0.507]\\
2007 &[ -0.030, 0.276]&[ 0.055, 0.361]&[ 0.018, 0.432]&[ -0.045, 0.324]&[ -0.042, 0.491]\\
2008 &[ -0.065, 0.296]&[ 0.072, 0.359]&[ 0.016, 0.462]&[ -0.011, 0.312]&[ -0.010, 0.477]\\
2009 &[ -0.026, 0.281]&[ 0.052, 0.381]&[ -0.025, 0.435]&[ 0.027, 0.386]&[ 0.049, 0.468]\\
2010 &[ -0.004, 0.240]&[ 0.089, 0.397]&[ 0.030, 0.444]&[ -0.004, 0.307]&[ -0.054, 0.390]\\
2011 &[ 0.000, 0.265]&[ 0.027, 0.348]&[ 0.031, 0.404]&[ 0.002, 0.293]&[ 0.011, 0.387]\\
2012 &[ 0.015, 0.243]&[ 0.029, 0.386]&[ 0.028, 0.412]&[ -0.025, 0.250]&[ -0.001, 0.397]\\
2013 &[ -0.026, 0.208]&[ 0.005, 0.372]&[ 0.015, 0.402]&[ -0.075, 0.229]&[ -0.006, 0.431]\\
\hline
\hline
\end{tabular}
}
\caption{Bounds on ATE of Type 1 with CLR adjustment under basic Assumptions}\label{table:T1EstCLR}
\end{table}

\begin{table}
  \centering
   \scalebox{.7}{
\begin{tabular}{cccccc}
\hline
\hline
Covariate Group & 1 &2 &3 &4 &5 \\ \hline
Year &[ LB, UB]&[ LB, UB]&[ LB, UB]&[ LB, UB]&[ LB, UB]\\
\hline
1976 &[ -0.304, 0.778]&[ -0.441, 0.745]&[ -0.399, 0.869]&[ -0.062, 1.152]&[ -0.177, 1.107]\\
1977 &[ -0.461, 0.733]&[ -0.527, 0.672]&[ -0.283, 0.941]&[ -0.036, 1.094]&[ -0.245, 1.133]\\
1978 &[ -0.585, 0.631]&[ -0.472, 0.721]&[ -0.192, 1.108]&[ -0.098, 1.216]&[ -0.111, 1.230]\\
1979 &[ -0.559, 0.607]&[ -0.594, 0.686]&[ -0.304, 1.123]&[ 0.085,  1.518]&[ -0.236, 1.191]\\
1980 &[ -0.501, 0.732]&[ -0.687, 0.715]&[ -0.318, 1.157]&[ -0.019, 1.538]&[ -0.071, 1.303]\\
1981 &[ -0.415, 0.744]&[ -0.660, 0.720]&[ -0.644, 0.947]&[ 0.015, 1.497]&[ -0.190, 1.148]\\
1982 &[ -0.448, 0.758]&[ -0.696, 0.728]&[ -0.431, 1.213]&[ -0.065, 1.594]&[ -0.388, 1.513]\\
1983 &[ -0.306, 0.765]&[ -0.854, 0.712]&[ -0.559, 1.172]&[ 1.481, 3.698]&[ -0.231, 1.714]\\
1984 &[ -0.365, 0.748]&[ -0.600, 0.838]&[ -0.359, 1.340]&[ 0.618, 2.639]&[ -0.375, 1.630]\\
1985 &[ -0.472, 0.756]&[ -0.828, 0.814]&[ -0.337, 1.339]&[ -0.263, 1.509]&[ -0.304, 1.445]\\
1986 &[ -0.448, 0.781]&[ -0.726, 0.945]&[ -0.288, 1.396]&[ -0.018, 1.725]&[ -0.418, 1.467]\\
1987 &[ -0.351, 0.876]&[ -1.188, 0.621]&[ -0.290, 1.496]&[ 1.817, 4.019]&[ -0.167, 1.546]\\
1988 &[ -0.476, 0.692]&[ -0.749, 0.933]&[ -0.434, 1.195]&[ 0.040, 1.919]&[ 0.026, 1.848]\\
1989 &[ -0.649, 0.671]&[ -0.709, 0.822]&[ -0.346, 1.556]&[ -0.241, 1.729]&[ -0.048, 1.686]\\
1990 &[ -0.490, 0.849]&[ -0.745, 0.915]&[ -0.343, 1.458]&[ 0.153, 1.933]&[ -0.193, 1.719]\\
1991 &[ -0.522, 0.720]&[ -0.920, 0.767]&[ -0.149, 2.012]&[ -0.124, 1.568]&[ -0.110, 1.717]\\
1992 &[ -0.551, 0.749]&[ -0.790, 0.844]&[ -0.513, 1.278]&[ -0.245, 1.463]&[ -0.134, 1.788]\\
1993 &[ -0.525, 0.746]&[ -0.703, 0.958]&[ -0.280, 1.563]&[ -0.172, 1.408]&[ 0.062, 1.849]\\
1994 &[ -0.575, 0.792]&[ -0.704, 0.868]&[ -0.218, 1.624]&[ 0.567, 2.426]&[ -0.104, 1.847]\\
1995 &[ -0.397, 0.993]&[ -0.741, 0.900]&[ -0.078, 1.889]&[ -0.353, 1.685]&[ -0.302, 1.540]\\
1996 &[ -0.584, 0.834]&[ -0.835, 0.880]&[ -0.212, 2.136]&[ 0.129, 2.045]&[ 0.173, 2.406]\\
1997 &[ -0.475, 1.053]&[ -0.754, 0.941]&[ -0.044, 1.978]&[ -0.274, 1.662]&[ -0.233, 1.958]\\
1998 &[ -0.934, 0.932]&[ -0.892, 1.112]&[ 0.195, 2.362]&[ -0.047, 1.881]&[ 0.074, 2.256]\\
1999 &[ -1.430, 0.718]&[ -1.186, 0.857]&[ -0.309, 2.022]&[ -0.147, 1.722]&[ -0.124, 2.040]\\
2000 &[ -0.708, 1.227]&[ -0.783, 1.233]&[ 0.263, 2.372]&[ -0.587, 1.638]&[ 0.280, 2.341]\\
2001 &[ -0.931, 1.285]&[ -1.173, 0.753]&[ -0.079, 2.033]&[ -0.242, 1.909]&[ -0.171, 1.984]\\
2002 &[ -1.184, 1.109]&[ -1.090, 1.147]&[ -0.544, 1.212]&[ -0.026, 1.929]&[ -0.339, 1.829]\\
2003 &[ -0.849, 1.167]&[ -0.822, 0.972]&[ -0.307, 1.484]&[ -0.054, 1.781]&[ -0.148, 1.830]\\
2004 &[ -0.862, 0.878]&[ -1.059, 0.853]&[ -0.373, 1.420]&[ -0.088, 1.749]&[ 0.173, 2.217]\\
2005 &[ -0.873, 1.034]&[ -0.856, 0.819]&[ -0.517, 1.468]&[ 0.010, 1.938]&[ -0.232, 1.990]\\
2006 &[ -0.774, 1.250]&[ -1.107, 0.782]&[ -0.655, 1.220]&[ 0.133, 2.237]&[ -0.148, 2.106]\\
2007 &[ -1.408, 0.711]&[ -1.367, 0.714]&[ -0.375, 1.645]&[ 0.077, 2.167]&[ -0.056, 2.181]\\
2008 &[ -1.238, 0.708]&[ -1.669, 0.507]&[ -0.482, 1.385]&[ 0.020, 2.149]&[ -0.102, 2.266]\\
2009 &[ -1.129, 0.825]&[ -1.252, 0.886]&[ -0.259, 1.644]&[ -0.528, 1.479]&[ 0.125, 2.854]\\
2010 &[ -1.108, 1.385]&[ -1.854, 0.592]&[ -0.547, 1.492]&[ 0.003, 2.341]&[ 0.519, 3.157]\\
2011 &[ -1.393, 0.989]&[ -0.970, 1.122]&[ -0.538, 1.553]&[ 0.068, 2.360]&[ 0.787, 3.573]\\
2012 &[ -2.520, 0.408]&[ -1.125, 0.871]&[ -0.340, 2.115]&[ 1.277, 3.821]&[ 0.875, 3.757]\\
2013 &[ -1.583, 1.445]&[ -0.972, 0.961]&[ -0.165, 2.362]&[ 0.790, 3.170]&[ 0.854, 3.824]\\
\hline
\hline
\end{tabular}
}
\caption{Bounds on ATE of Type 2 with CLR adjustment under basic Assumptions}\label{table:T2EstCLR}
\end{table}

\end{document}